\documentclass[showpacs,twocolumn,amsmath,amssymb,floatfix,aps,prb,footinbib,superscriptaddress]{revtex4-1}
\usepackage{graphicx}
\usepackage{psfrag}
\usepackage{amsmath,amsfonts}
\usepackage{color}

\def\up{\uparrow}									
\def\down{\downarrow}							
\def\+{\dagger}							

\renewcommand\vec{\mathbf}		

\newcommand{\pint}{\mathcal{P}\!\!\!\!\!\!\int\!\!}

\newcommand{\df}{\partial_{\epsilon}\hspace*{0.1em} f}

\newcommand{\ep}{\hat{e}_p}
\newcommand{\dtep}{\partial_t \ep}

\newcommand{\pol}{P}

\newcommand{\gn}{\Gamma_{\text{N}}}
\newcommand{\gf}{\Gamma_{\text{F}}}
\newcommand{\g}{\Gamma}

\definecolor{blassgelb}{rgb}{1.0,1.0,0.9}
\definecolor{red}{rgb}{1.0,0.0,0.0}

\definecolor{green}{rgb}{1.0,0.0,0.0}

\begin{document}
\title{Theory of spin pumping through an interacting quantum dot tunnel coupled to a ferromagnet with time-dependent magnetization}

\author{Nina Winkler}
\affiliation{Theoretische Physik, Universit\"at Duisburg-Essen and CENIDE, 47048 Duisburg, Germany}
\author{Michele Governale}
\affiliation{School of Chemical and Physical Sciences and MacDiarmid Institute for Advanced Materials and Nanotechnology, Victoria University of Wellington, PO Box 600, Wellington 6140, New Zealand}
\author{J\"urgen K\"onig}
\affiliation{Theoretische Physik, Universit\"at Duisburg-Essen and CENIDE, 47048 Duisburg, Germany}

\date{\today}
\begin{abstract}
We investigate two schemes for  pumping spin adiabatically from a ferromagnet through an interacting quantum dot into a normal lead, which exploit the possibility to vary in time the ferromagnet's magnetization, either its amplitude or its direction.
For this purpose, we extend a diagrammatic real-time technique for pumping to situations in which the leads' properties are time dependent. 
In the first scheme, the time-dependent magnetization amplitude is combined with a time-dependent level position of the quantum dot to establish both a charge and a spin current.
The second scheme uses a uniform rotation of the ferromagnet's magnetization direction to generate a pure spin current without a charge current. 
We discuss the influence of an interaction-induced exchange field on the pumping characteristics.

\end{abstract}



\maketitle

\section{Introduction}

The last decades have seen intense research in the field of spintronics, a branch of physics and electrical engineering whose aim  is to combine electrical and magnetic functionalities in the same solid-state system by exploiting the spin degree of freedom of the charge carriers.~\cite{,zutic_spintronics_2004,awschalom_challenges_2007,sin-nmat-2012}
In particular, mechanisms to generate and control spin currents have been the object of several theoretical and experimental investigations.  
A subset of these studies, which is particularly relevant to the present paper, focused on quantum-dot spin valves, that is devices made up of a quantum dot sandwiched between two  ferromagnetic leads. Experimentally, these types of systems have been realised with self-assembled  InAs quantum dots,\cite{hamaya_spin_2007,hamaya_electric_2007,hamaya_kondo_2007,hamaya-prb-2008,hamaya-apl-2008} small metallic grains,\cite{deshmukh-2002,bernand-mantel-2006,mitani-2008,bernand-mantel-2009,birk-2010}  nanowires,\cite{hofstetter-2010} nanotubes \cite{sahoo-2005,jensen-2004,hauptmann-2008} and molecular systems. \cite{pasupathy_kondo_2004} Quantum dots contacted with ferromagnetic leads  have also attracted  a considerable theoretical interest. \cite{bulka-2000,rudzinski-2001,sergueev-2002,zhang-2002,lopez-2003,koenig_interaction_2003,martinek_kondo_qd_2003,martinek_kondo_rgm_2003,bulka-2003,choi-2004,cottet-prl-2004,braun_theory_2004,braig-2005,utsumi-prb-2005,fransson-epl-2005,pedersen-2005,weymann-zero-2005,weymann-tunnel-2005,weymann_cotunneling_2005,braun_frequency_2006,cottet-prb-2006,simon-prb-2007,matsubayashi-2007,splettstoesser_adiabatic_2008,lindebaum_spin_2009,schenke-2009,sothmann_transport_2010,koller-2012}

A possible mechanism to generate a pure spin current, that is a spin current without an associated charge current, relies on a ferromagnet with a rotating magnetization. This mechanism is the corner stone of recent spin-battery proposals.~\cite{tserkovnyak_enhanced_2002,brataas_spin_2002,tserkovnyak_spin_2002,mahfouzi_microwave_2010,watson_experimental_2003,costache_electrical_2006}
In a related study, it has been predicted that the rotation of the magnetization direction of a magnetic quantum dot  that is sandwiched between a ferromagnet and a normal lead gives rise to charge pumping.~\cite{bender_tserkovnyak_brataas_2010}. Similarly, spin pumps using time-dependent magnetic fields acting on a molecule have been proposed.~\cite{fransson1,fransson2}. 
These time-dependent transport problems 
can be formulated within the theory of charge and spin pumping in mesoscopic structures.~ 
\cite{buttiker_current_1994,brouwer_scattering_1998,aleiner-1998,zhou_mesoscopic_1999,moskalets-inelastic-2001,makhlin_counting_2001,moskalets_dissipation_2002,entin_adiabatic_2002,citro-2003,aono_adiabatic_2003,brouwer-2005,cota_ac_2005,splettstoesser_adiabatic_2005,sela_adiabatic_2006,splettstoesser_adiabatic_2006,fioretto-2008,arrachea_relation_2006,splettstoesser_adiabatic_2008,braun-burkard,winkler_diagrammatic_2009,cavaliere_nonadiabatic_2009,hernandez-2009}
However, pumping usually refers to the situation when the properties of the conductor and not those of the leads (as it is the case for the spin battery) are varied in time. 
The regime of adiabatic pumping is realised when the period of the time variation is long compared to the characteristic dwell time of the carriers in the system. This regime is particularly interesting  from the theoretical point of view since it leads to an understanding of the non-equlibrium caused by the explicit time dependence of the system without introducing the additional complications associated with higher-order terms in the frequency of the time-dependent parameters.

\begin{figure}
\includegraphics[width=0.3\textwidth,angle=0]{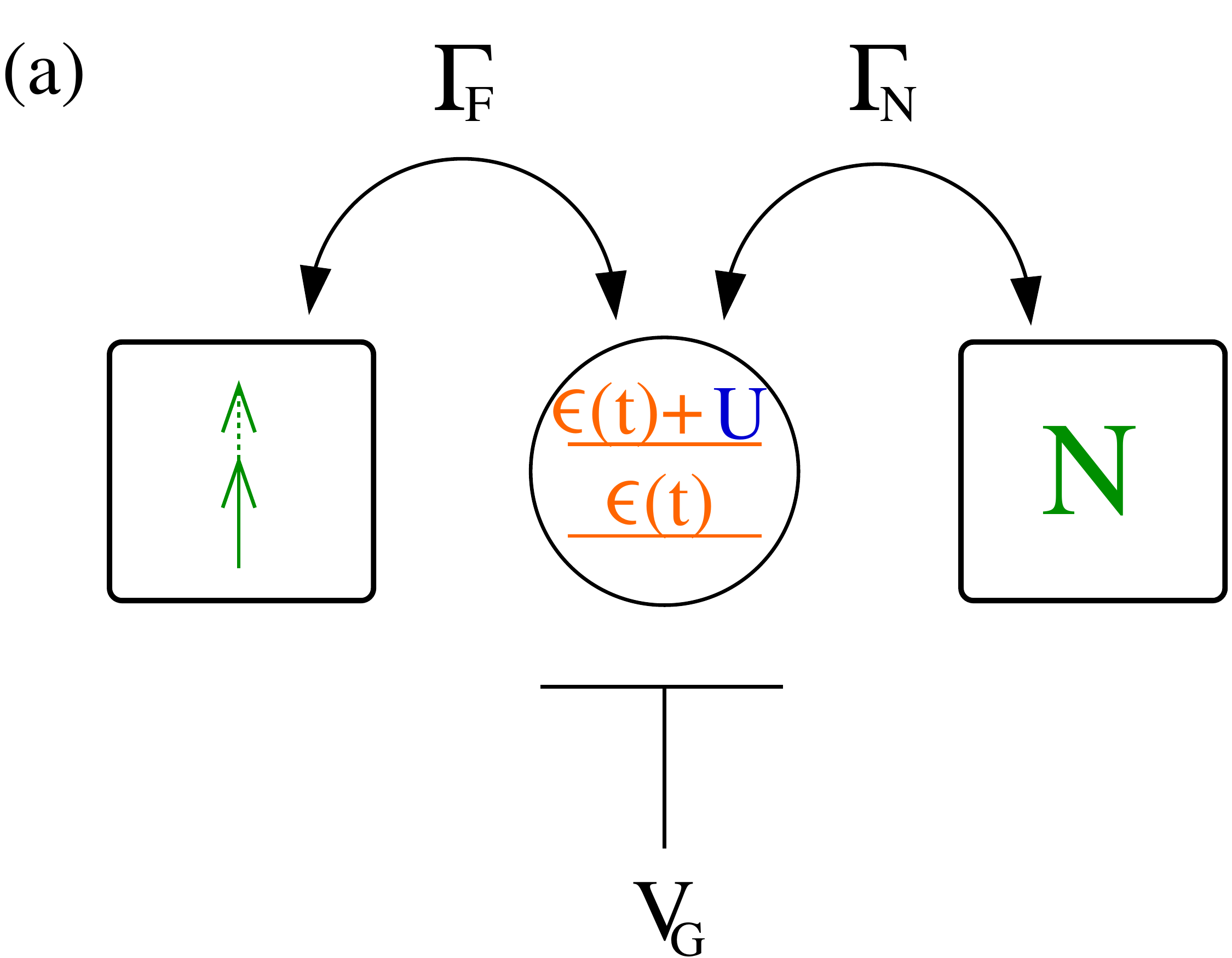}\\
~\\ 	\includegraphics[width=0.3\textwidth,angle=0]{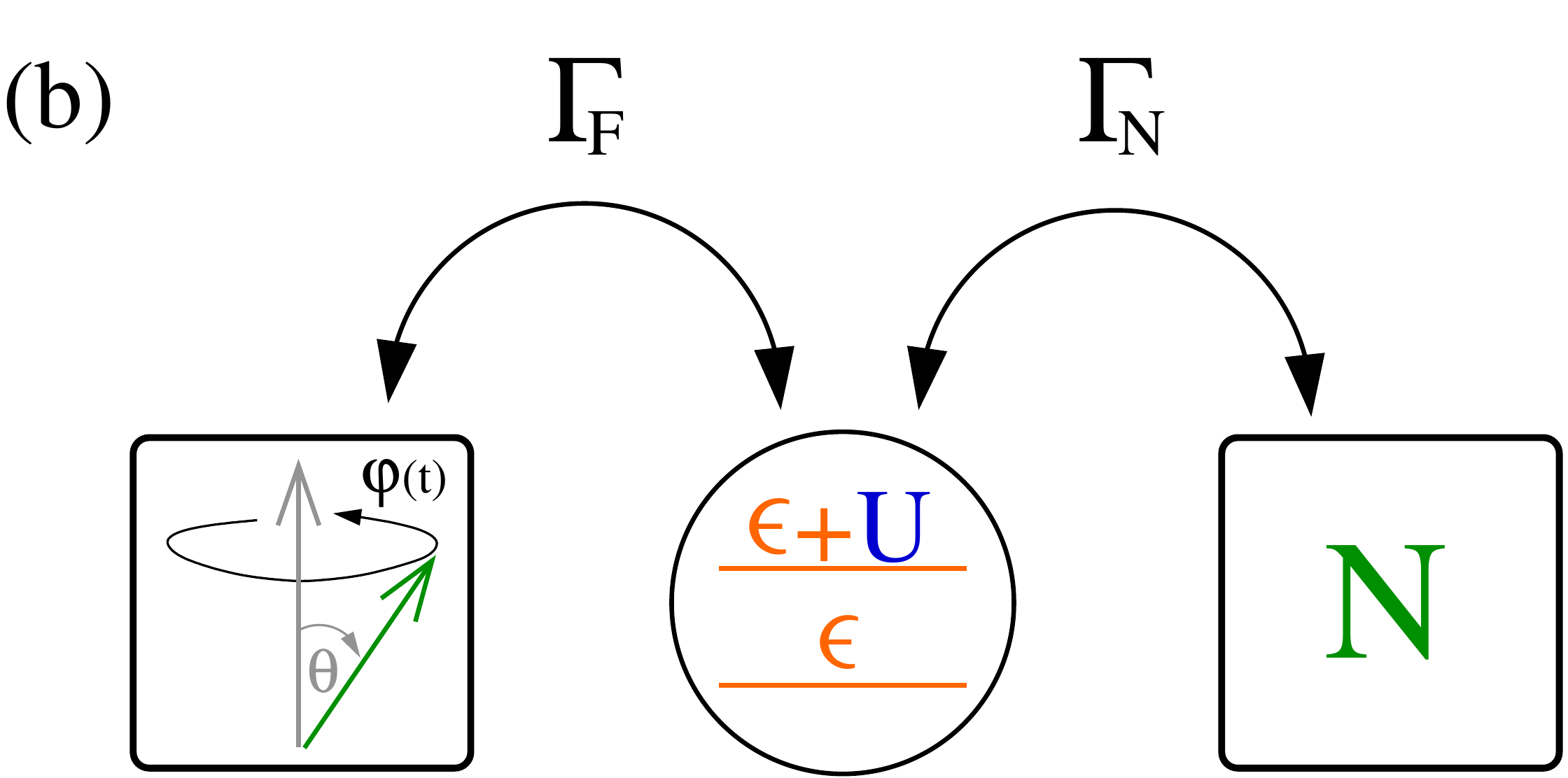}
	\caption{(Color online) Sketch of the F-dot-N structure. It consists of a quantum dot, tunnel coupled to a ferromagnetic and a non-magnetic lead, which are kept at the same chemical potential.\\ 
	(a) Schematic description of pumping with the amplitude of the magnetization of the ferromagnetic lead and the level-position $\epsilon$ of the dot.\\
	(b) Schematic description of pumping with the x- and y-component of the magnetization of the ferromagnetic lead.
	}
	\label{fig_model-1dot}
\end{figure}

In this paper we focus on a spin battery realised in a structure  composed  of  a quantum dot, tunnel coupled to a ferromagnetic and a non-magnetic lead. The spin-battery operation is due to the time-dependent magnetization of the ferromagnetic lead. 
We employ a diagrammatic real-time theory for adiabatic pumping through quantum dots with ferromagnetic leads~\cite{splettstoesser_adiabatic_2006,splettstoesser_adiabatic_2008} that we extend to account for time-dependent magnetizations. This approach consists in a systematic perturbative expansion in powers of the tunnel-coupling strengths and of the pumping frequency, while  treating the on-site Coulomb interaction on the quantum dot exactly. We consider two different pumping schemes: First, we choose the amplitude of the magnetization of the ferromagnetic lead and the level position of the dot as pumping parameters. Second, we pump by changing periodically  the direction of the magnetization. 
A sketch of the system under investigation and of the two pumping schemes is shown in Fig.~\ref{fig_model-1dot}.
 
The main results we find are the following.
In the first pumping scheme the pumped spin current is, in general, accompanied with a finite charge current. 
There are, however, special values of the system parameters, for which the pumped charge current turns out to be zero due to a cancellation of two counteracting contributions to the overall pumped charge. 
As a consequence, in the first pumping scheme the generation of a pure spin current requires a fine tuning of the system parameters .  
This is profoundly different to the second pumping scheme in which the pumped charge vanishes for any choice of the system parameters. 
We are able to derive a compact analytical formula for the pumped spin current and to discuss its dependence on the gate voltage, the ratio of the tunnel couplings to the ferromagnet and the normal lead, and the exchange field that acts on the quantum dot spin as a consequence of a spin-dependent tunnel coupling of the dot level to the ferromagnet.

This paper is structured as follows. 
After introducing the model of the system in Sec.~\ref{sec_model}, we extend in Sec.~\ref{formalism} the diagrammatic real-time technique to calculate the pumped charge and spin through systems including ferromagnets with time-dependent magnetization.
The diagrammatic rules resulting from this are given in Sec.~\ref{sec_rules}.  The main part of the paper is Sec.~\ref{sec_results} where the results for both pumping scheme are presented. We conclude by summarizing our results in Sec.~\ref{sec_conclusion}. 

To keep all formulas transparent we set $\hbar =1$ throughout the paper. 

\section{Model}
\label{sec_model}

The Hamiltonian of the system reads  $H= H_{\text{dot}}+H_{\text{lead N}}+H_{\text{lead F}}+H_{\text{tun}}$. 
The single-level quantum dot is described by the Anderson impurity model
%
%
\begin{equation}
\label{eq_H_dot}
	H_{\text{dot}}=\sum\limits_{\sigma} \epsilon\, d^\+_{\sigma}\, d_{\sigma} + U\, n_\up	\, n_\down \,,
\end{equation}
where $\epsilon$  is the  spin-degenerate energy level of the dot, $U$ the on-site Coulomb interaction, $d_{\sigma}$ ($d^{\dagger}_{\sigma}$) the annihilation (creation) operator of an electron on the dot with spin $\sigma=\up,\down$, and $n_{\sigma}=d^\+_{\sigma} d_{\sigma}$ the corresponding number operator. 
The Hilbert space for the quantum dot is four-dimensional.The basis states are denoted by $\left| 0\right\rangle$, $\left| \up \right\rangle$, $\left| \down \right\rangle$, and $\left| \rm{d}\right\rangle=d_\uparrow^\dagger d_\downarrow^\dagger |0\rangle$ corresponding, respectively,  to the dot being empty, singly occupied with spin  $\sigma=\up,\down$, or doubly occupied. 

The leads are modeled by reservoirs of non-interacting electrons (kept at the same chemical potential $\mu$ which we set to zero) with Hamiltonians 
%
%
\begin{align}
\label{eq_H_leads}
	H_{\text{lead N}}&=\sum\limits_{k,\sigma} \epsilon _{k}\,c^\+_{k\sigma}\,c_{k\sigma}  \\ 
	H_{\rm{lead\, F}}&=\sum\limits_{k,\alpha} E_{k \alpha}a^\dagger_{k \alpha}a_{k \alpha} \,, 
\end{align}
where $c_{k\sigma }$  ($c^{\dagger}_{k\sigma }$) is the annihilation (creation) operator for an electron with spin $\sigma$ and wave vector $k$ in the normal lead, while $a_{k\alpha }$ ($a^{\dagger}_{k\alpha }$) is the annihilation (creation) operator of an electron with majority/minority spin $\alpha=\pm$ and wave vector $k$ in the ferromagnetic lead. 
The density of states $\varrho_{\rm{N}}(\omega)=\sum_k \delta(\omega-\epsilon_k)$ in the normal lead is independent of spin.
The ferromagnet, on the other hand, is described by majority- and minority-spin bands, $E_{k\alpha} = \varepsilon_k+E_\alpha$, that are shifted relative to each other by a finite Stoner splitting $\Delta E \equiv E_{k -} - E_{k +}$.
This leads to a spin-dependent density of states $\varrho_{\rm{F},-}(\omega)= \varrho_{\rm{F},+}(\omega-\Delta E)$.
The degree of spin polarization at energy $\omega$ is characterized by $P(\omega) = \left[\varrho_{\rm{F},+}(\omega) -\varrho_{\rm{F},-}(\omega)  \right]/\left[\varrho_{\rm{F},+}(\omega) +\varrho_{\rm{F},-}(\omega)  \right]$.
We remark that the majority and minority spin direction in the ferromagnet (denoted by $\alpha=\pm$) may, in general, be different from the spin quantization axis (with $\sigma=\uparrow,\downarrow$) that we choose for the quantum dot and the normal lead.
The corresponding Fermi operators are connected via the transformation $a_{k\alpha}^\dagger = \sum_\sigma A_{\alpha \sigma} a_{k\sigma}^\dagger$ with
\begin{equation}
\label{maj/min}
	\left( A_{\alpha \sigma} \right) = \left( 
	\begin{array}{cc}
	e^{- i \varphi/2 } \cos (\theta/2) & e^{i \varphi/2 } \sin(\theta/2)
	\\ 
	- e^{- i \varphi/2 } \sin(\theta/2) & e^{i \varphi/2 } \cos(\theta/2) 
	\end{array}
	\right)	
	\, ,
\end{equation} 
where the polar angle $\theta$ and azimuthal angle $\varphi$ define the ferromagnet's magnetization direction $\hat{e}_p= \left( \sin\theta \, \cos\varphi , \sin\theta \, \sin\varphi , \cos\theta \right)^\text{T}$ in the (time-independent) coordinate system with the $z$-axis chosen along the spin-quantization axis of quantum dot and normal lead.

The tunnel coupling between dot and leads is described by the spin-conserving tunneling Hamiltonian
%
%
\begin{equation}
\label{eq_H_tun}
	H_{\text{tun}}=\sum\limits_{k,\sigma}\left( V_{\rm{N}}\; c^\+_{k\sigma}\,d_{\sigma} +V_{\rm{F}}\, a^\+_{k\sigma}	\,d_{\sigma} +\text{h.c.}\right)\,,
\end{equation}
where $V_{\rm{N}}$ and $V_{\rm{F}}$ are the energy- and spin-independent tunnel-matrix elements.
Tunneling introduces a finite lifetime of the dot states, characterized by the intrinsic line width 
$\Gamma_{\rm{N}}(\omega) =2\pi \varrho_{\rm{N}}(\omega)  |V_{\rm{N}}|^2$ 
and
$\Gamma_{\rm{F},\alpha}(\omega) =2\pi \varrho_{\rm{F},\alpha}(\omega)  |V_{\rm{F}}|^2$. 
For later use, we define $\gf(\omega)=\left[\Gamma_{\rm{F},+}(\omega) +\Gamma_{\rm{F},-}(\omega) \right] / 2$, which implies $P(\omega)\,\gf(\omega)=\left[\Gamma_{\rm{F},+}(\omega) -\Gamma_{\rm{F},-}(\omega)  \right] /2$,
or, equivalently, $\Gamma_{\rm{F},\alpha}(\omega) = \Gamma_{\rm{F}}(\omega) [1+\alpha\, P(\omega)]$. 
Finally, we define the total intrinsic linewidth $\Gamma(\omega) = \Gamma_\text{F}(\omega) + \Gamma_\text{N}(\omega)$.

The main goal of this paper is to investigate charge and spin pumping schemes relying on the variation of the ferromagnet's magnetization $\vec{M}(t)=M(t) {\hat{e}}_p(t)$, specifically its {\em amplitude} $M(t)$ or {\em direction} ${\hat{e}}_p(t)$, in time.
The variation of the magnetization {\em amplitude} can be microscopically modeled by time-dependent shifts of the majority and minority bands, $E_\alpha(t)$, which leads to a time-dependent Stoner splitting $\Delta E(t)$.
The variation of the magnetization {\em direction}, on the other hand, is described by time-dependent angles $\theta(t)$ and $\varphi(t)$ in Eq.~(\ref{maj/min}), which leads to an explicit time dependence of the operators $a_{k\pm (t)}$.

In order to achieve pumping in the adiabatic regime, two system parameters need to be varied in time with a relative phase. 
For a time-dependent magnetization direction, these could be the x- and y-component of the polarization, $P_x(t)=P \sin \theta (t) \cos \varphi (t)$ and $P_y(t)=P \sin \theta   (t) \sin \varphi (t)$.
If, on the other hand, the magnetization direction is fixed and only its amplitude is varied via a time-dependent Stoner splitting $\Delta E(t)$, a second pumping parameter is needed.
Therefore, in the following derivation, we allow for a time-dependent level position $\epsilon \left(t \right) $, which can be experimentally controlled by a gate voltage.

\section{Formalism}
\label{formalism}

The problem under consideration is, in general, complicated since it combines a few interacting degrees of freedoms of the quantum dot with a large number of non-interacting degrees of freedom in the leads and an explicit time dependence.
For a time-independent Hamiltonian, it is possible to integrate out the lead degrees of freedom to obtain an effective description of the system in terms of the reduced density matrix.
Then one can perform a diagrammatic expansion of the time evolution of the reduced density matrix to write the kinetic equations of the reduced system.\cite{koenig_resonant_1996,koenig_zero_1996}

In the presence of an explicit time dependence of the Hamiltonians describing the dot and/or the tunnel coupling, the kinetic equations can still be formally derived but they become much more complicated, such that a full solution is only achievable in special cases.\cite{cavaliere_nonadiabatic_2009} 
In the limit of adiabatic pumping, however, the kinetic equations can be simplified considerably by performing an adiabatic expansion, i.e., an expansion in the pumping frequency $\Omega$ (or powers in time derivatives of the pumping parameters), assuming that the response of the system is much faster than the change of the system parameters.\cite{splettstoesser_adiabatic_2006,splettstoesser_adiabatic_2008}

In the present context, though, also the lead Hamiltonian acquires a time dependence.
This possibility is not included in the diagrammatic technique presented in Ref.~\onlinecite{splettstoesser_adiabatic_2006} where a central step relies on integrating out the lead degrees of freedom, for which a time-independent equilibrium distribution is assumed.
In order to treat time-dependent lead Hamiltonians, we start by performing an adiabatic expansion already at the very first step for the Hamiltonian before integrating out the lead degrees of freedom.
The expansion should be performed about the time $t$ at which the charge and spin currents are calculated. 
At any different time $\tau$, the Hamiltonian $H(\tau) = H_0(t) + V(\tau)$ is approximated by
\begin{eqnarray}
\label{H_0}
	H_0(t) &=& H_{\text{dot}}(t) + H_{\text{lead N}} + H_{\text{lead F}}(t)
\\
\label{V}
  	V(\tau) &=& H_{\text{tun}} + (\tau-t) \left[ \dot{H}_{\text{dot}}(t) + \dot{H}_{\text{lead F}}(t)\right] 
\, .
\end{eqnarray}
Higher time derivatives of $H_{\text{dot}}$ and $H_{\text{lead F}}$ are neglected in the adiabatic expansion.
The time derivative of the dot Hamiltonian is given by
\begin{equation}
	\dot H_{\text{dot}}(t) = \dot \epsilon \sum\limits_{\sigma} \, d^\+_{\sigma}\, d_{\sigma} \, .
\end{equation}
To derive the time derivative of the ferromagnetic lead, we use $\dot a_{k\sigma} = 0$ (in Schr\"odinger picture) to obtain $\dot a_{k\alpha}^\dagger
= \alpha (\dot \theta/2) a_{k\overline{\alpha}}^\dagger {- i} (\dot{\varphi}/2) \left( \alpha \cos\theta \, a_{k\alpha}^\dagger - \sin\theta \, a_{k\overline{\alpha}}^\dagger \right)$, where we introduced the notation $\overline{\alpha} = -\alpha$.
This yields
\begin{eqnarray}
	\dot{H}_{\text{lead F}}(t) &=& \sum\limits_{k,\alpha} \dot E_{\alpha} \; a^\dagger_{k \alpha}a_{k \alpha} 
	\nonumber \\
	&+& \frac{\Delta E}{2} \sum\limits_{k,\alpha}
  	\left(-\dot \theta + i \alpha \dot{\varphi} \sin\theta \right) 
	a^\+_{k \alpha} a_{k \overline{\alpha}} 
	\, , 
\end{eqnarray}
which includes non-spin-flip (diagonal) terms in the first and spin-flip (off-diagonal) ones on the second line.

We remark that there is no time derivative of the tunnel Hamiltonian since it does not depend explicitly on time, which is evident from Eq.~(\ref{eq_H_tun}).
For integrating out the lead degrees of freedom, however, it is convenient to rewrite in the following the tunnelling  Hamiltonian in terms of the majority/minority electron operators by using the time-dependent transformation Eq.~(\ref{maj/min}). 

After having separated the Hamiltonian $H(\tau)$ into the time-independent part $H_0(t)$ of the decoupled system and the correction $V(\tau)$ due to tunneling and the explicit time dependence of the system parameters, we can 
continue in a similar way as in the derivation of the diagrammatic rules for a time-independent system.\cite{koenig_resonant_1996, koenig_zero_1996}
First, we express the quantum-statistical expectation value of any operator $A$ at time $t$ as integral over the Keldysh contour $K$, that runs from time $\tau=-\infty$ to $t$ and then back to $-\infty$,
\begin{equation}
\label{average-A}
	\langle A(t)\rangle = \text{tr} \left[ \varrho_0 T_K \exp \left( -i \int_K d\tau V(\tau)_I \right) A(t)_I \right] \, .
\end{equation}
The Keldysh time ordering operator $T_K$ orders all operators along the Keldysh contour, the index $I$ indicates interaction picture with respect to $H_0(t)$, and $\rho_0$ is the (full) density matrix at time $-\infty$.
The latter is assumed to be a tensor product of the density matrices for the quantum dot and the leads, with the leads being in an equilibrium state determined by $H_{\text{lead N}}$ and $H_{\text{lead F}}(t)$, respectively. 
By doing so, we neglect nonequilibrium distributions of the ferromagnet.
This is justified as long as the time scale for spin-relaxation towards equilibrium in the ferromagnet is much shorter than the time scale for transport, given by $\hbar/\Gamma$.
In that case, the non-adiabatic corrections of the ferromagnet's density matrix are negligible in comparison to the non-adiabatic corrections of the transport processes, and we only need to develop a theoretical description of the latter.
The next step is to expand the exponential in powers of $V(\tau)_I$.    
In the diagrammatic language, each $V(\tau)_I$ is represented by a vertex.
In the present context, there are three different types of vertices: $H_{\text{tun}}$ (represented by a full circle) contains one dot and one lead operator, $\dot H_{\text{dot}}(t)$ (represented by an empty circle) two dot operators, and $\dot H_{\text{lead F}}(t)$ (represented by a double cross) two lead operators.

At this stage, it is possible to perform the partial trace over the lead degrees of freedom.
By using Wick's theorem, we contract the lead operators in pairs. 
In the diagrams, each contraction is represented by a tunneling line.
The diagram for the full time evolution of the reduced density matrix from $-\infty$ to $t$ is then a sequence of irreducible blocks, defined as the parts of a diagram, where a vertical line at any time $\tau$ crosses at least one tunneling line.   
This infinite series of irreducible blocks can be summed up by a Dyson equation.

To obtain the kinetic equation for the matrix elements $p^{\chi_1}_{\chi_2} (t)=\langle \chi_1|\varrho_{\text{red}}(t)|\chi_2\rangle$ of the reduced density matrix $\varrho_{\text{red}}$, we use for the operator $A$ in Eq.~(\ref{average-A}), the projector $ |\chi_2\rangle \langle \chi_1|$ with $\chi_1,\chi_2 \in \{0, \up, \down, \text{d}\}$. 
This leads to
%
%
\begin{equation}
\label{eq_master}
	\dfrac{d}{dt}\mathbf{p}\left( t\right) = \int\limits_{-\infty}^{t}dt'\;\mathbf{W}\left( t,t'\right)\mathbf{p}	\left( t'\right) \,,
\end{equation}
where $\mathbf{p}$ is the vector of all relevant density matrix elements, $\mathbf{p}=(p_0,p_\uparrow,p_\downarrow,p_d,p^\downarrow_\uparrow,p^\uparrow_\downarrow)^\text{T}$.
The diagonal matrix element $p_\chi \equiv p^\chi_\chi$ is the occupation probability of  the state $\chi$ of the dot, while the off-diagonal ones, $p^\downarrow_\uparrow = (p^\uparrow_\downarrow)^*$ describe coherent superpositions of up- and down-spins.
The kernel $\mathbf{W}\left( t,t'\right)$ is a $6\times 6$ matrix which acts on the vector $\mathbf{p}(t')$.
It represents an irreducible block in the diagrammatic language and describes transitions between the states at time $t'$ and time $t$.

In order to distinguish the charge and the spin degrees of freedom, it is convenient to parametrize the reduced density matrix not by the six elements of the vector $\mathbf{p}$ but rather use the probability vector $\mathbf{P} = \left( P_0, P_1, P_d  \right)^\text{T} =\left( p_0, p_\uparrow + p_\downarrow, p_d \right)^\text{T}$ and the spin vector $\mathbf{S} = \left( S_x, S_y, S_z \right)^\text{T} 
= \left( \frac{p^\uparrow_\downarrow + p^\downarrow_\uparrow}{2}, i \; \frac{p^\uparrow_\downarrow - p^\downarrow_\uparrow}{2}, \frac{p_\uparrow - p_\downarrow}{2} \right)^\text{T} $ instead.
In this basis, the kinetic equations take the more intuitive form
%
%
\begin{subequations}
\begin{align}
\label{eq_master_P}
	\dfrac{d}{dt}\mathbf{P}\left( t\right) 
	&= \!\! \int\limits_{-\infty}^{t}\!\!\!dt'\;\mathbf{W}_p \left( t,t'\right)\mathbf{P}\left( t'\right) 
	+\mathbf{W}_s \left( t,t'\right)\mathbf{S}	\left( t'\right)  \\
\label{eq_master_S}
	\dfrac{d}{dt}\mathbf{S}\left( t\right) 
	&= \!\! \int\limits_{-\infty}^{t}\!\!\!dt'\;\mathbf{M}_p \left( t,t'\right)\mathbf{P}\left( t'\right) 
	+\mathbf{M}_s \left( t,t'\right)\mathbf{S}	\left( t'\right) .
\end{align}
\end{subequations}
The matrix elements of the $3 \times 3$ matrices $\mathbf{W}_p\left( t,t'\right)$, $\mathbf{W}_s\left( t,t'\right)$, $\mathbf{M}_p\left( t,t'\right)$, and $\mathbf{M}_s\left( t,t'\right)$ are the proper linear combinations of the matrix elements of the $6 \times 6$ matrix $\mathbf{W}\left( t,t'\right)$.
The first contribution to Eq.~(\ref{eq_master_S}), which depends on $\mathbf{P}$, can be interpreted as spin accumulation, while the second one describes relaxation and coherent rotation of the spin $\mathbf{S}$.

We remark here that spin rotational symmetry about a given axis simplifies the structure of the kinetic equations Eqs.~(\ref{eq_master_P}) and (\ref{eq_master_S}). 
The simplified form of the kinetic equations is derived in Appendix~\ref{appendix_general_trafo}.
This symmetry applies when the ferromagnet's magnetization direction $\hat e_p$ is constant in time.
But even for a time-dependent $\hat e_p(t)$, the spin rotational symmetry is present for the instantaneous part of the kinetic equations (defined in the next section).
  
\subsection{Expansion of the kinetic equation}

Although we have already linearized the time dependence of the Hamiltonian $H(\tau)$ about the time $t$, we still need to perform a systematic adiabatic expansion for the kinetic equations.
For this task, we follow Ref.~\onlinecite{splettstoesser_adiabatic_2006}.
The lowest ({\em instantaneous}) order describes the equilibrium situation when all parameters are frozen to their values at time $t$,
\begin{align}
	0=\mathbf{W}_t^{(\text{i})}\;\mathbf{p}_t^{(\text{i})}   \,.
\end{align}
The next-order ({\em adiabatic}) correction contains all contributions linear in the pumping frequency i.e., with one first-oder time derivative appearing,
\begin{align}
 	\dfrac{d}{dt}\;\mathbf{p}_t^{(i)}
	&=\;\mathbf{W}_t^{(\text{i})}\mathbf{p}_t^{(\text{a})} +\;\mathbf{W}_t^{(\text{a})}\mathbf{p}_t^{(\text{i})} 
	+\;\partial \mathbf{W}_t^{(\text{i})}\;\dfrac{d}{dt}\;\mathbf{p}_t^{(\text{i})} \,.
\end{align}
Here, the index $t$ indicates the time with respect to which the adiabatic expansion has been performed and the superscript indicates the order of the adiabatic expansion. 
By construction, $\mathbf{W}_t^{(\text{a})}$ contains exactly one empty circle or one double cross vertex representing
$\dot H_{\text{dot}}(t)$ or $\dot H_{\text{lead F}}(t)$, respectively.
Furthermore, we have introduced the Laplace transform to define the abbreviation $\mathbf{W}_t^{(\text{i/a})}
=\left.\mathbf{W}_t^{(\text{i/a})}\left(z\right)\right|_{z=0_+} =\int_{-\infty}^{t} dt'\,\mathbf{W}_t^{(\text{i/a})}(t-t')$ 
as well as $\partial \mathbf{W}_t^{(\text{i})} = \left.\partial \mathbf{W}_t^{(\text{i})}\left(z\right)/\partial z \right|_{z=0_+}$.

On top of the adiabatic expansion, we perform a systematic perturbation expansion in powers of the tunnel-coupling strengths $\Gamma$.  The order of the expansion in the tunnel coupling will be indicated by an integer superscript. For example, $\mathbf{W}^{(\text{a},1)}_t$ indicates the first order in $\Gamma$ of the adiabatic correction of the kernel. This expansion is straightforward and can be easily performed (details can be found  in Ref.~\onlinecite{splettstoesser_adiabatic_2006}.)

The perturbation expansion of the kernel starts in first order in $\Gamma$, which corresponds to sequential tunneling processes described by diagrams containing one tunneling line.
The perturbation expansion of the instantaneous probability vector starts in zeroth order in $\Gamma$, since it corresponds to the time-independent problem with all parameters frozen at time $t$. Thus, the normalization conditions read $\mathbf{e}^{\text{T}} \, \mathbf{P}^{(\text{i},0)}_t =1$ and  $\mathbf{e}^{\text{T}} \, \mathbf{P}^{(\text{i},1)}_t =0$ with $\mathbf{e}^\text{T}=(1,1,1)^\text{T}$. 
The adiabatic correction (i.e. first order in pumping frequency $\omega$) to the probabilities starts in minus first order in $\Gamma$ and obeys the normalization conditions $\mathbf{e}^{\text{T}} \, \mathbf{P}^{(\text{a},-1)}_t =0$ and $\mathbf{e}^{\text{T}} \, \mathbf{P}^{(\text{a},0)}_t =0$.  
When going to the limit of weak tunnel coupling, $\Gamma \rightarrow 0$, one needs to keep in mind that the adiabaticity condition, $\Omega \ll \Gamma$, has to remain fulfilled. 
Therefore, $\mathbf{p}^{(\text{a},-1)}_t \propto \Omega/ \Gamma$ does not diverge.

\subsection{Pumped charge and spin current}
 
We are interested in the pumped charge and the pumped spin (projected along a time-independent spin quantization axis) through the quantum dot.
The pumped charge and spin currents flowing into the normal-metal lead can be written, respectively,  as
\begin{subequations}
\begin{align}
	I_{\text{N}}\left(t\right)&=e\int_{-\infty}^{t}dt'\mathbf{e}^{\text{T}} \,
	\mathbf{W}^{\text{N},Q}\left(t,t'\right)\mathbf{p}\left(t'\right) \\
	\mathbf{J}_{\text{N}}(t)&=\dfrac{1}{2} \int\limits_{-\infty}^{t}dt' \mathbf{e}^{\text{T}} \,\mathbf{W}^{\text{N},\mathbf{S}}\left(t,t'\right) \mathbf{p}\left(t'\right)
\end{align}
\end{subequations}
where $\mathbf{W}_t^{\text{N},Q} \left(t,t'\right) =  \mathbf{W}_t^{\text{N},\uparrow} \left(t,t'\right) + \mathbf{W}_t^{\text{N},\downarrow} \left(t,t'\right)$, 
and the matrix elements of the kernel $\mathbf{W}^{\text{N},\sigma}_t \left(t,t'\right)$ are evaluated with the same diagrammatic rules as for $\mathbf{W}_t \left(t,t'\right)$ but with the difference that each diagram is multiplied with the number of electrons with spin $\sigma$ entering  the normal lead during the transition minus the ones leaving the normal lead.
The $z$-component of the vector $\mathbf{W}^{\text{N},\mathbf{S}}= (\mathbf{W}^{\text{N},S_x},\mathbf{W}^{\text{N},S_y},\mathbf{W}^{\text{N},S_z})$ is given by $\mathbf{W}^{\text{N},S_z}= \mathbf{W}_t^{\text{N},\uparrow} \left(t,t'\right) - \mathbf{W}_t^{\text{N},\downarrow} \left(t,t'\right)$, and the $x$- and $y$-component are obtained by the same expression but with the spin quantization axis being chosen along the $x$- and $y$-direction, respectively.

In the same way as for the kinetic equation, we perform an adiabatic expansion of the charge and spin currents. 
The instantaneous currents vanish since, in the absence of a bias voltage the instantaneous terms describe the equilibrium situation.

By integrating the pumped charge and spin currents over one pumping cycle  $\mathcal{T} = \frac{2 \pi}{\Omega}$ we obtain the pumped charge and spin as
%
%
\begin{subequations}
\begin{align}
\label{charge}
	Q_{X} & =  \int \limits _0^{\mathcal{T}} \; I_{\text{N}}(t) \; dt \\
\label{spin}
\mathbf{S}_{X} 
& =  \int \limits _0^{\mathcal{T}} \; 
\mathbf{J}_{\text{N}}(t) \; dt  \,,
\end{align}
\end{subequations}
where the index $X$ indicates the pumping parameters. 

\section{Diagrammatic Rules}
\label{sec_rules}

We explain here how to evaluate the instantaneous diagrams $\mathbf{W}_t^{(\text{i})}
=\left.\mathbf{W}_t^{(\text{i})}\left(z\right)\right|_{z=0_+}$ and $\partial \mathbf{W}_t^{(\text{i})} = \left.\partial \mathbf{W}_t^{(\text{i})}\left(z\right)/\partial z \right|_{z=0_+}$, as well as the first adiabatic correction $\mathbf{W}_t^{(\text{a})}$. 
 
\subsection{Rules for the instantaneous diagrams}

In the following, we write down the rules for the instantaneous diagrams $\mathbf{W}_t^{(\text{i},n)}(z)$ to $n$-th order in the tunnel coupling.
An example of a diagram belonging to the instantaneous kernel element $(\mathbf{W}_t^{(\text{i},1)})_{0\up}$ and its adiabatic correction $(\mathbf{W}_t^{(\text{a},1)})_{0\up}$ 
are shown in Fig.~\ref{fig_example_diagrams}. All diagrams contributing to these matrix elements are presented and evaluated in Appendix~\ref{appendix_example_diagrams}.
%
%
\begin{figure}
	\includegraphics[height=0.21\textwidth,angle=0]{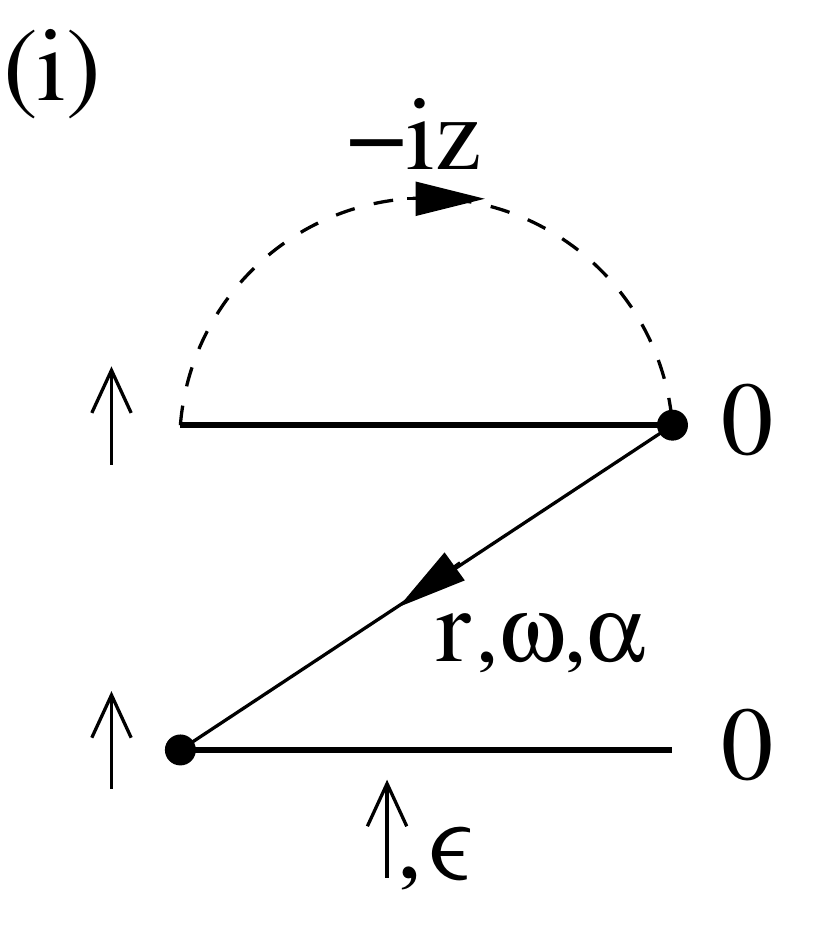}
	\includegraphics[height=0.21\textwidth,angle=0,clip]{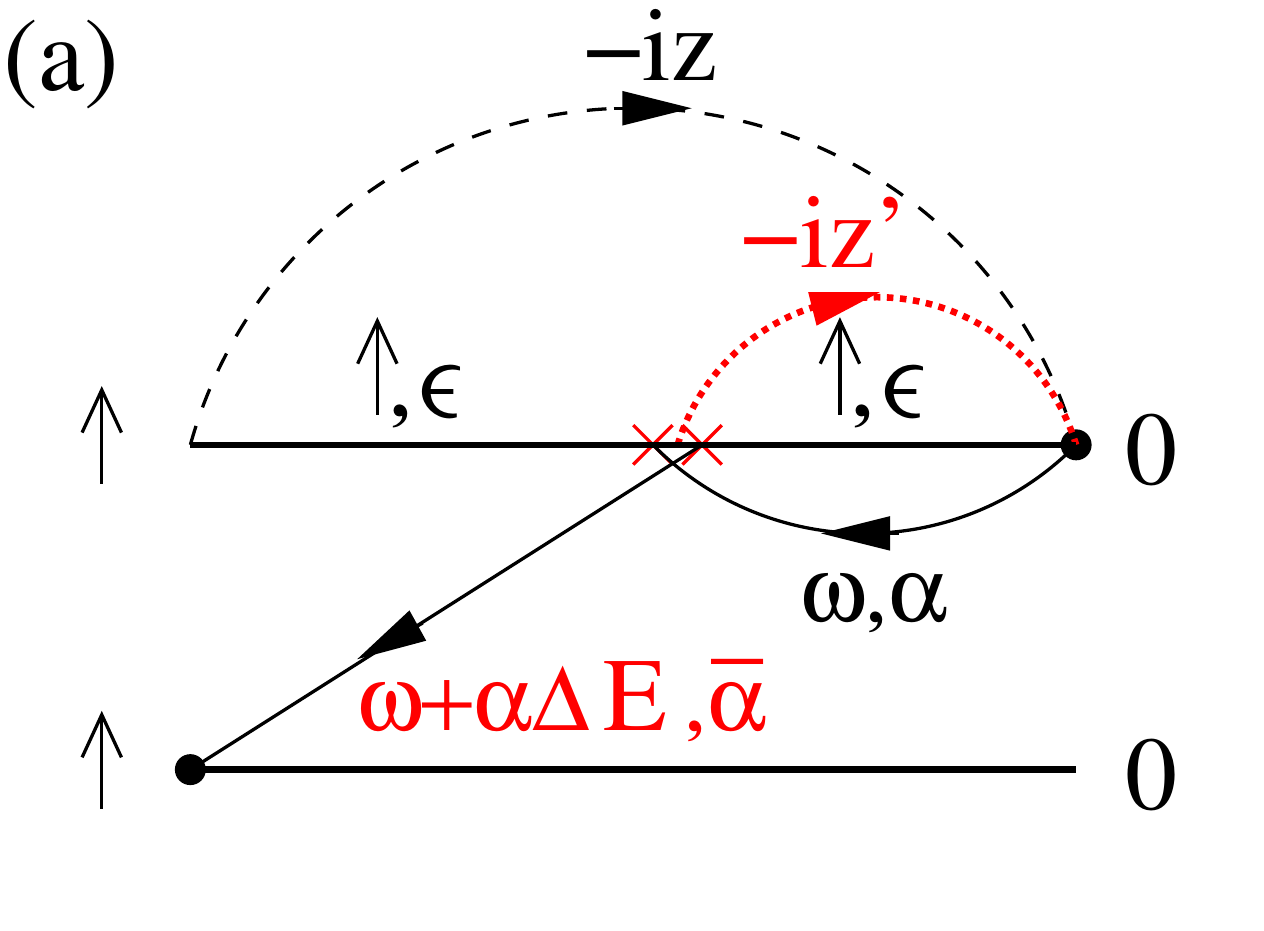}
	\caption{Examples of diagrams belonging to the instantaneous (i) and adiabatic (a) kernel in first order in the tunnel-coupling strength~$\g$, respectively.
	}
	\label{fig_example_diagrams}
\end{figure}
\begin{enumerate}
\item[(1)] Draw all topologically different diagrams with $ 2n $ full-circle vertices connected in pairs by directed tunneling lines. Assign a reservoir index $r= \text{N},\text{F}$, an energy $ \omega $, and a spin index ($ \sigma =\up,\down $ for $r=\text{N}$  and $ \alpha = \pm $ for $r=\text{F}$) to each of these lines. Assign quantum-dot states $ \chi \in \{ 0,\up,\down,\text{d} \}$ and the corresponding energies $ E_{\chi} \in \{ 0, \epsilon,\epsilon, 2\epsilon+U \}$ to each element of the Keldysh contour between two vertices. Furthermore, draw an external line with the (imaginary) energy $ -iz $ from the upper leftmost beginning of a dot propagator to the upper rightmost end of a dot propagator. 
\item[(2)] For each time segment between two adjacent vertices (independent on whether they are on the same or on opposite branches of the Keldysh contour) write a resolvent $ 1/R $, where $ R $ is the difference of left-going minus right-going energies (including energies of tunneling lines and the external line - the positive imaginary part of $ iz $ will keep all resolvents regularized).
\item[(3)] The contribution of a tunneling line consists of a prefactor $1/(2\pi)$ and a factor $ \gn (\omega) $ for $r=\text{N}$ or $ \Gamma_{\text{F},\alpha}(\omega) $ for $r=\text{F}$.
This is multiplied with $ f^+(\omega) \equiv f(\omega) $ if the tunneling is going backwards with respect to the Keldysh contour and $ f^-(\omega) \equiv 1-f(\omega)$ if it is going forward.
Here, $f(\omega)$ is the Fermi function and 
$\Gamma_{\rm{F},\alpha}(\omega) = \Gamma_{\rm{F}}(\omega) [1+\alpha\, P(\omega)]$.
\item[(4)] Each full-circle vertex with an incoming or outgoing tunneling line (with spin $\sigma$ or $\alpha$ for 
$r=\text{N}$ or $r=\text{F}$, respectively) changes the dot state assigned to the Keldysh contour by adding or removing an electron with spin $\sigma$.
In the case $r=\text{F}$, the matching of the different spin quantization axes of dot and lead is achieved introducing a 
prefactor $A_{\alpha \sigma}$ for a vertex with an outgoing line and $A_{\sigma \alpha}=\left(A_{\alpha \sigma}\right)^*$ for a vertex with an incoming line.
\item[(5)] The overall prefactor is given by $ (-i)(-1)^{b+c} $ where $ b $ is the total number of vertices on the backward propagator and $c$ the number of crossings of tunneling lines.
Furthermore, there is a minus sign for each vertex connecting dot states $|\up\rangle$ and $|\text{d}\rangle$. 
\item[(6)] Integrate over the energies of tunneling lines and sum over the reservoirs and the spin indices of the leads. 
\end{enumerate}

\subsection{Rules for the adiabatic diagrams}

The adiabatic corrections to the kernels are described by diagrams that contain one vertex that is associated with the time derivative of the Hamiltonian.
This may be an empty-circle vertex for $(\tau-t)\dot H_{\text{dot}}(t)$ or a double-cross vertex for $(\tau-t)\dot H_{\text{lead F}}(t)$, respectively.
No tunneling line is attached to the empty-circle vertex since $\dot H_{\text{dot}}(t)$ does not contain any lead operator.
In contrast, there are two lead operators in $\dot H_{\text{lead F}}(t)$.
As a consequence, two tunneling lines are coupled to a double-cross vertex: first (with respect to the Keldysh contour) an incoming and then an outgoing one.

When performing Wick's theorem, the two lead operators of the double cross may be contracted either with each other or with two other lead operators of full-circle vertices.
The earlier possibility, however, does not contribute (any diagram with such a self-contracted double cross vertex on the upper propagator cancels with the diagram obtained from it by moving the double-cross vertex to the lower propagator.)

To evaluate the diagrams $\mathbf{W}_t^{(\text{a},n)}(z)$ the rules for $\mathbf{W}_t^{(\text{i},n)}(z)$ need to be modified in the following way:

\begin{itemize}
\item [(1')] In addition to the $2n$ full circle vertices there is either one open-circle or one double-cross vertex which, in principle, can sit everywhere on the contour.
The latter is connected to two tunneling lines of the ferromagnet: first (with respect to the Keldysh contour) an incoming and then an outgoing line. 
The double-cross vertex may be either diagonal or off-diagonal in spin.
In the first case, the two lines carry the same energy $\omega$ and the same spin $\alpha$.
In the second case,
 one carries $\omega$ and $\alpha$, and the other one $\omega+\alpha \Delta E$ and $\overline{\alpha}$ ($\equiv -\alpha$).
Add an external frequency line with (imaginary) energy $ -iz'$ from the empty-circle or double-cross vertex to the upper right corner of the diagram. 
\item [(3')] When applying rule (3) for the two tunneling lines connected to a double-cross vertex, only one prefactor 
$\Gamma_{\text{F},\alpha}(\omega) /(2\pi)$ (the one of the line which carries $\omega$ and $\alpha$) has to be taken into account.
\item [(4')] An empty-circle vertex comes with a factor $\dot E_\chi(t)$, where $E_\chi(t)$ is the energy of the dot state $\chi$ assigned to the Keldysh contour segment where the empty circle vertex is placed.
For a double-cross vertex that is diagonal in spin, the factor is $\dot E_\alpha(t)$.
The factor for a double-cross vertex off-diagonal in spin is $(\Delta E/2) \left( -\dot \theta + i \alpha \dot \varphi \sin \theta \right)$, where $\alpha$ is the spin of the outgoing line.
Furthermore, one needs to perform a first derivative with respect to $z'$ and then send $z'$ to $0^+$.
\item [(5')] When applying rule (5), $b$ is the total number of all vertices (including full circle, empty circle, and double cross vertices).
Similarly, the number $c$ includes also crossings of tunneling lines at a double cross vertex.
\end{itemize}
 
We remark that the contributions of diagrams with the rightmost vertex being a double cross cancel out since for each such diagram with the double cross on the upper propagator, there is a partner diagram with the double cross on the lower propagator which only differs by a minus sign due to rule (5).
 
Furthermore, we remark that for time-independent magnetization direction of the ferromagnet, $\dot \theta =\dot \varphi = 0$, only double-cross vertices diagonal in spin space appear.
In this case, it is possible to directly integrate in time over all positions of the empty-circle or double-cross vertex between two full-circle vertices. 
As a consequence one does not need to explicitly draw this empty-circle or double-cross vertex but the adiabatic correction to the kernels are obtained from the same diagrams as for the instantaneous ones with modified rules. 
This route has been used in Ref.~\onlinecite{splettstoesser_adiabatic_2006} to account for a time dependence of the dot level position.

\section{Results}
\label{sec_results}

We start by deriving the expressions for the charge and spin currents to lowest-order in the tunnel coupling strength.
To this order, only the instantaneous kernels $\mathbf{W}_t^{(i,1)}$ to first order in $\Gamma$ enter both the instantaneous limit and the adiabatic correction of the kinetic equations,
\begin{align}
\label{kin_i}
	\mathbf{0}&=\mathbf{W}_t^{(\text{i},1)}\;\mathbf{p}_t^{(\text{i},0)}   
 \\
\label{kin_a}
  	\dfrac{d}{dt}\;\mathbf{p}_t^{(\text{i},0)}
	&=\;\mathbf{W}_t^{(\text{i},1)}\; \mathbf{p}_t^{(\text{a},-1)}   \,.
\end{align}
As a consequence, the kinetic equations and the expressions for the pumped charge and spin current to this order are independent of the pumping scheme, i.e., the choice of pumping parameters. 
Furthermore, there is rotational spin symmetry about the axis $\ep$, i.e., the kinetic equations take the form derived in Appendix~\ref{appendix_general_trafo}.

After explicit calculations, that are summarized in Appendix~\ref{appendix_lowest_order}, we obtain
%
%
\begin{subequations}
\begin{align}
\label{ccur_a_0}
	I_{\rm{N}}^{(a,0)}(t) &= -\,e \;\;\dfrac{\frac{\Gamma_{\rm{N}}}{\Gamma}}{1-\pol^2 \, \frac{\Gamma_{\rm{F}}^2}{\Gamma^2} }\;\;\dfrac{d}{d t}\langle n\rangle^{(i,0)} \\
\label{scur_a_0}
\mathbf{J}_{\text{N}}^{(\text{a},0)}(t)
&= \frac{\Gamma_{\text{N}} }{2\Gamma}  
  \frac{P\frac{\Gamma_{\rm{F}}}{\Gamma}}{1-\pol^2 \, \frac{\Gamma_{\rm{F}}^2}{\Gamma^2} } \; \ep\; \dfrac{d}{d t}\langle n\rangle^{(i,0)}
\end{align}
\end{subequations}
for the charge and the spin current. 
The latter is polarized along $\ep$.
Both currents have a similar dependence on $P$, $\Gamma_{\rm{N}}$, $\Gamma_{\rm{F}}$, and the instantaneous part of the average number of electrons $\langle n\rangle=P_1+2P_d$ expanded to zeroth order in $\Gamma$.
As a consequence, for each moment in time, the ratio between the component of the spin current $J_{\text{N}}^{(\text{a},0)}(t) = \mathbf{J}_{\text{N}}^{(\text{a},0)}(t) \cdot \ep$ along the (instantaneous) symmetry axis and the charge current is given by
%
%
\begin{equation}
	\frac{J_{\rm{N}}^{(a,0)}(t)}{I_{\rm{N}}^{(a,0)}(t)} = - \, \dfrac{1}{2e} \pol \, \dfrac{\Gamma_{\rm{F}}}{\Gamma} \,.
\end{equation}

In order to calculate the pumped charge and spin per cycle, we need to specify the pumping scheme.
As announced in the introduction, we will consider two different scenarios. 

\subsection{Pumping scheme A: time-dependent magnetization amplitude}

In pumping scheme A, we assume that the magnetization amplitude $M(t)$ of the ferromagnet changes in time while its directions remains fixed. 
Experimentally, this could be realized by using paramagnetic diluted magnetic semiconductors which exhibit a large Zeeman splitting. 
A small, externally applied magnetic field would, then, spin polarize the lead, and the degree of spin polarization could be varied in time by making the external field time dependent.

For pumping scheme A, it is convenient to choose the same spin quantization axis for the dot and the ferromagnet, $+=\up$ and $-=\down$. 
The variation of the magnetization amplitude can be microscopically modeled by time-dependent Stoner splitting 
$\Delta E=E_{k-} - E_{k+}$.

To be more specific, the majority- and minority-spin bands are shifted relative to each other in time, $E_\pm(t)$, in such a way that both the Fermi energy and the total number of electrons in the ferromagnet remain constant.  
This is, at low temperature, fulfilled for $\dot{E}_+/\dot{E}_- = - \varrho_{\text{F},-} (\epsilon_F) /\varrho_{\text{F},+} (\epsilon_F)$.
As a consequence, both the polarization $\pol(t)$ and tunnel-coupling strength $\gf(t)$ vary in time.
This alone does not establish any adiabatic pumping since the time variations of $\pol(t)$ and $\gf(t)$ are in phase.
Therefore, we use the dot-level position $\epsilon$ as another out-of-phase time-dependent parameter. In this situation there are two pumping cycles occurring simultaneously: one cycle with $\{ \epsilon , \pol \}$ and another cycle with $\{ \epsilon , \gf \}$ as pumping parameters. 

In the following, we concentrate on the limit of weak pumping, i.e. we write the level position $\epsilon$ as well as the polarization $P$ or the tunnel-coupling strength $\Gamma_\text{F}$ as a sum of the  average value and a small variation, $\epsilon(t)=\bar{\epsilon} +\delta \epsilon(t)$, $\pol(t)=\bar{\pol} +\delta \pol(t)$, and $\gf(t)=\bar{\Gamma}_\text{F} +\delta \gf(t)$, and expand the pumped charge and spin to bilinear order in the variations.
Then, we integrate the charge and spin currents over one pumping cycle to obtain the pumped charge and spin, respectively.
The latter are proportional to the area of the pumping cycle in parameter space, quantified by the dimensionless quantities $\eta_1 =\int _0^{\mathcal{T}} dt \; \delta\beta\epsilon\; \delta \dot{\pol} $  and  $\eta_2 =\int _0^{\mathcal{T}} dt \; \delta\beta\epsilon\; \delta \dot{\Gamma}_\text{F}/\bar{\Gamma}_\text{F} $  for the cycles with $\{ \epsilon , \pol \}$ and $\{ \epsilon , \gf \}$, respectively. 

Before addressing the sum of the two pumping contributions, we analyze them separately.
The reason is that their relative weight to the total pumped charge and spin in the weak-pumping regime depends on the ratio $\nu=\bar{\pol} \, \eta_2/\eta_1$ that, in turn, depends on details of the ferromagnet's band structure.
To be specific it depends, at low temperature, on the density of states as well as the energy derivative of the density of states of the majority and minority spins at the Fermi energy.
Expressing them in terms of the polarization $P= P(\epsilon_\text{F})$ and the total density of states $\varrho_\text{F} = \varrho_\text{F}(\epsilon_\text{F}) = \frac{1}{2} [ \varrho_{\text{F},+} (\epsilon_\text{F})+ \varrho_{\text{F},-}(\epsilon_\text{F}) ]$ and their derivatives $P'= \partial P(\omega)/\partial \omega|_{\omega=\epsilon_\text{F}}$ and
$\varrho_\text{F}'= \partial \varrho_\text{F}(\omega)/\partial \omega|_{\omega=\epsilon_\text{F}}$ leads to
\begin{equation}
	\nu = \frac{ \frac{P^2 }{1-P^2 } \frac{P'}{P} } 
		{ \frac{\varrho_\text{F}' }{\varrho_\text{F} }  - 2 \frac{P^2}{1-P^2 } \frac{P' }{P}  } \,.
\end{equation}
For small polarizations $P$, the ratio $\nu$ scales as $P^2$ and is, thus, also small. 
This means that in this case the main contribution to pumping is due to the cycle with $\epsilon$ and $P$.
For large polarizations, $P\rightarrow 1$, on the other hand, $\nu\rightarrow -1/2$.
It is easy to show that for parabolic bands the following relation holds $-1/2 \le \nu \le 0$.

For arbitrary band structures, however, also positive values of $\nu$ are possible.
This can be seen, e.g., for a weak ferromagnet with a small Stoner splitting $\Delta E$ by expanding the ratio $\nu$ in $\Delta E$,
\begin{equation}
	\nu \approx \frac{1}{4} \Delta E^2 \left[ \frac{ \varrho''_\text{F} }{\varrho_\text{F} }  -
	 	\left( \frac{ \varrho'_\text{F} }{\varrho_\text{F} } \right)^2 \right] \, ,
\end{equation}
which is positive whenever $\varrho''_\text{F} / \varrho_\text{F} > (\varrho'_\text{F} / \varrho_\text{F})^2$.
For example, $\nu>0$ is realized  by a functional dependence of $\varrho_\text{F}(\omega) \propto \omega^{-\alpha}$ with $\alpha>0$.

\subsubsection{Contribution from pumping with $\epsilon$ and $P$}

First, we consider $\{ \epsilon , \pol \}$ as pumping parameters.
The  procedure described above yields for the 
pumped charge $Q_X$ and pumped spin $S_X=\mathbf{S}_X \cdot \hat{e}_p$ along the symmetry axis,

%
%
\begin{subequations}
\label{charge-spin}
\begin{align}
\label{}
	Q_{\epsilon, \pol} &= 2 e    \eta_1 
	  \dfrac{ \bar{\pol}  \, \frac{\bar{\Gamma}_{\rm{F}}^{2}}{\bar{\Gamma}^2}\, \frac{\bar{\Gamma}_{\rm{N}}}{\bar{\Gamma}} }	{\left( 1- \bar{\pol}^2 \, \frac{\bar{\Gamma}_{\rm{F}}^2}{\bar{\Gamma}^2} \right)^{2}} \; \dfrac{d}{d\beta	\bar{\epsilon}}   \left\langle \bar{n}\right\rangle ^{(\text{i},0)}  \\
\label{}
	S_{\epsilon, \pol} &=- \frac{1}{2}   \eta_1 
	  \dfrac{ \frac{\bar{\Gamma}_{\rm{F}}}{\bar{\Gamma}}\, \frac{\bar{\Gamma}_{\rm{N}}}{\bar{\Gamma}}  \left( 1+ \bar{\pol}^2 \, \frac{\bar{\Gamma}_{\rm{F}}^2}{\bar{\Gamma}^2} \right)}	{\left( 1- \bar{\pol}^2 \, \frac{\bar{\Gamma}_{\rm{F}}^2}{\bar{\Gamma}^2} \right)^{2}} \; \dfrac{d}{d\beta	\bar{\epsilon}}   \left\langle \bar{n}\right\rangle ^{(\text{i},0)}  ,
\end{align}
\end{subequations}
where $\eta_1 =\int _0^{\mathcal{T}} dt \; \delta\beta\epsilon\; \delta \dot{\pol} $ is the area of the pumping cycle in parameter space, $\left\langle \bar{n}\right\rangle ^{(\text{i},0)}$ the instantaneous average occupation number, where the level position has been replaced by its time average ${\bar{\epsilon}}$, and $\bar{\Gamma} =\bar{\Gamma}_\text{N} +\bar{\Gamma}_\text{F}$.
 
%
%
\begin{figure}
	\includegraphics[width=0.45\textwidth,angle=0,clip]{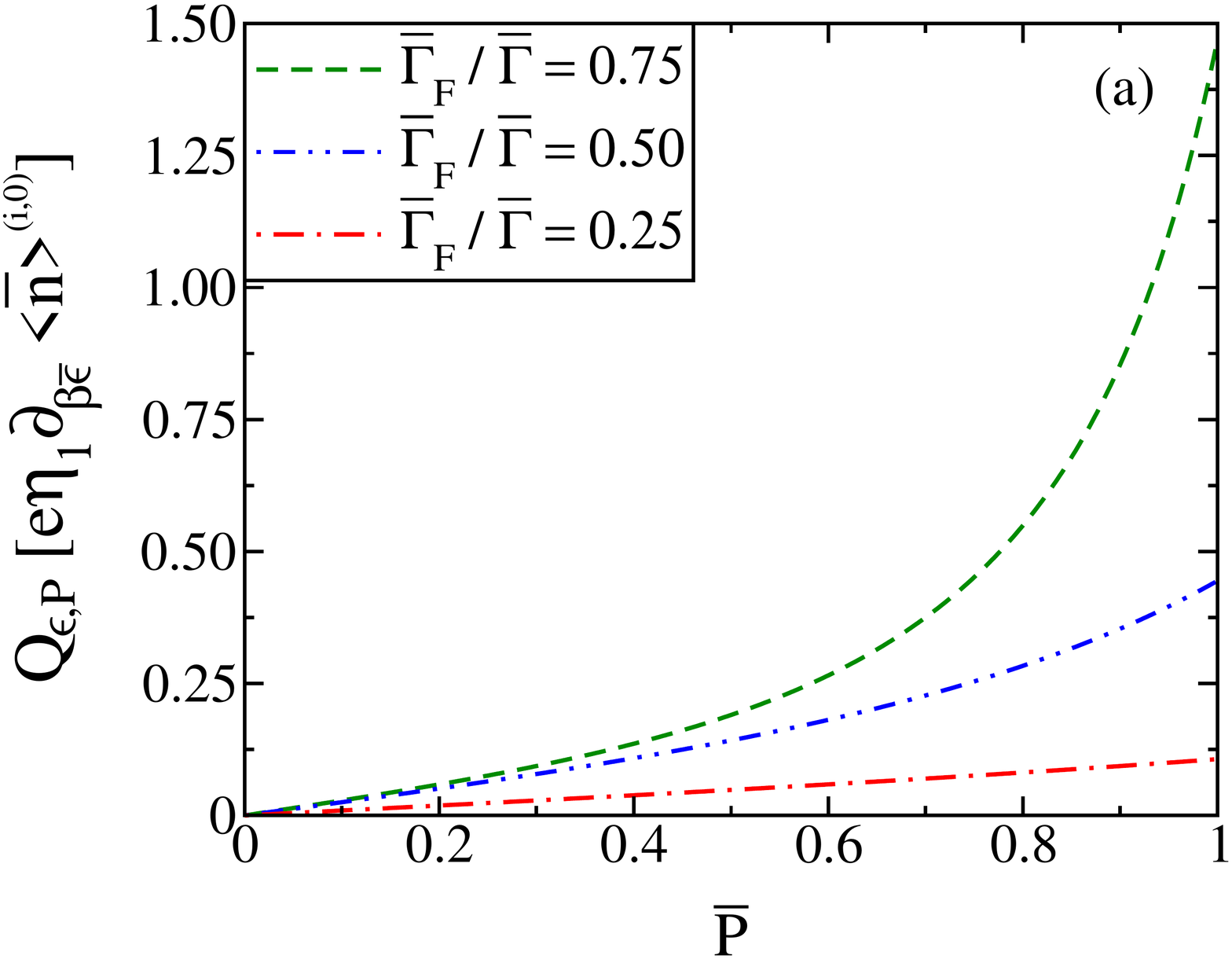} \\
	\includegraphics[width=0.45\textwidth,angle=0,clip]{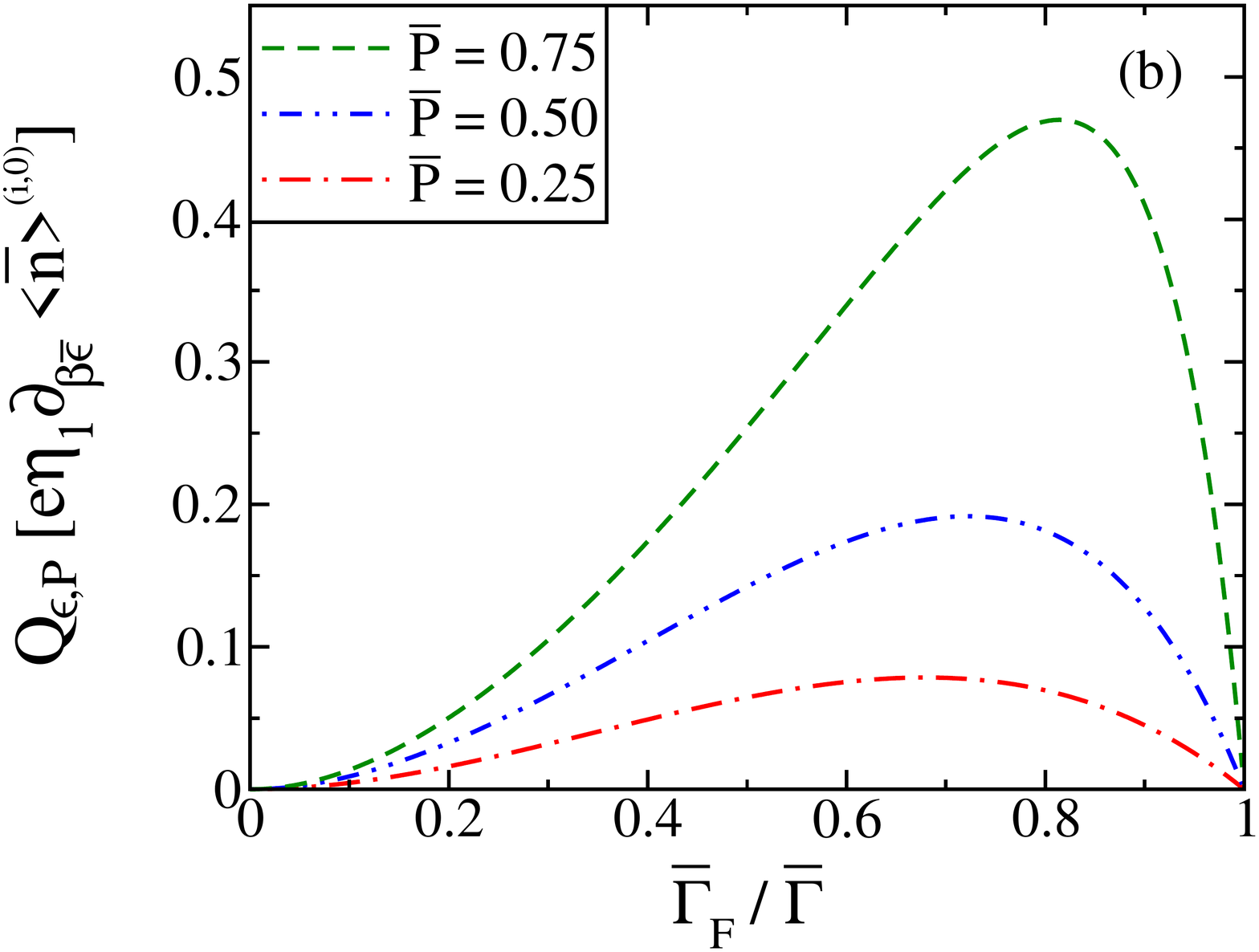}\\
	\caption{
	(Color online)  
	Pumped charge in units of $e \eta_1 \partial_{\beta \bar{\epsilon}}   \left\langle \bar{n}\right\rangle ^{(\text{i},0)}  $ 	as a function of  (a) the time average of the  lead polarization and (b) the relative tunnel-coupling strength  respectively. The pumping parameters are $\epsilon$ and $\pol$. }
	\label{fig_charge}
\end{figure}
%
%
\begin{figure}
	\includegraphics[width=0.45\textwidth,angle=0,clip]{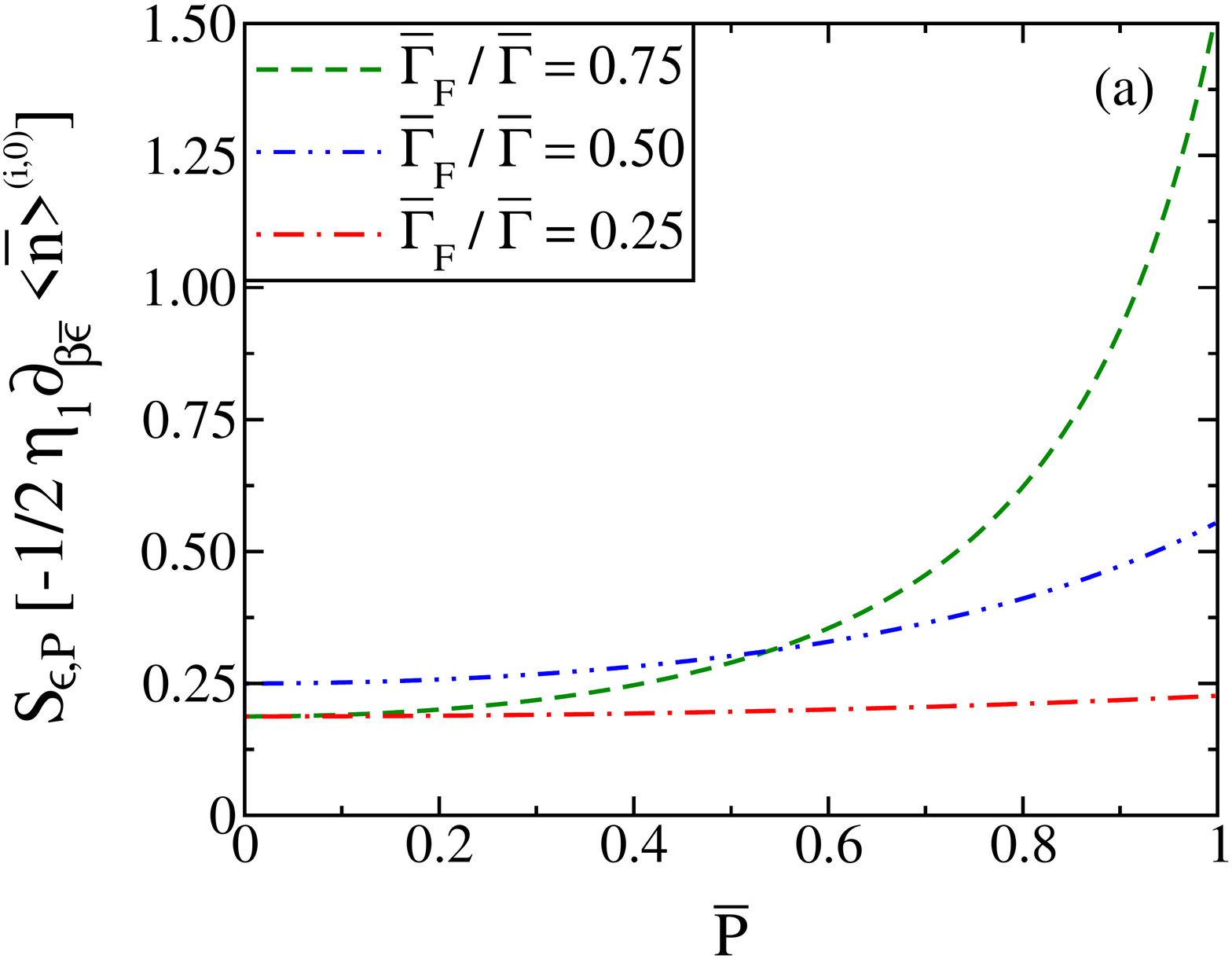} \\
	\includegraphics[width=0.45\textwidth,angle=0,clip]{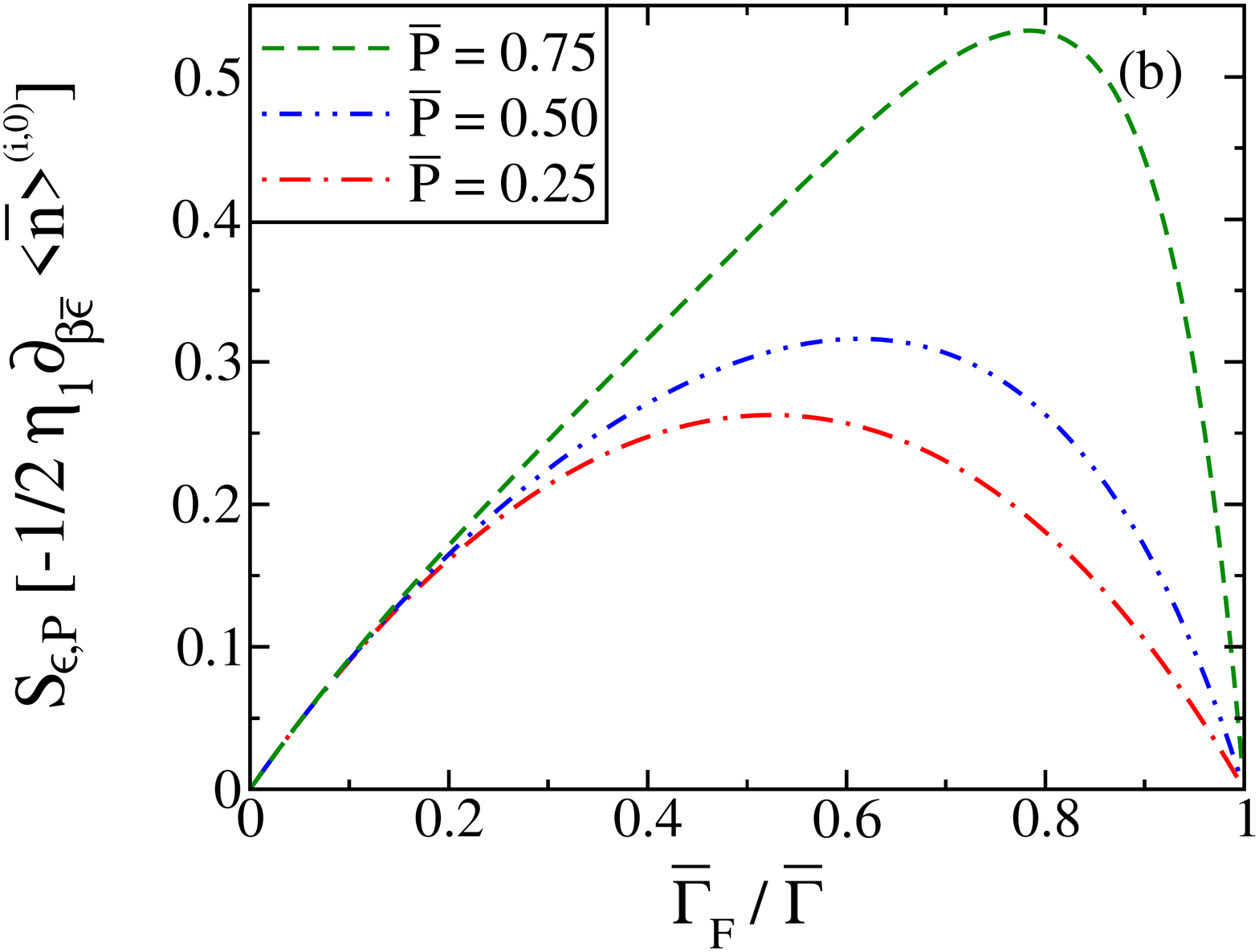}\\
	\caption{
	(Color online) 
	Pumped spin in units of $-\frac{1}{2} \eta_1 \partial_{\beta \bar{\epsilon}}   \left\langle \bar{n}\right\rangle ^{(\text{i},0)}  $ 	as a function of (a) the time average of the  lead polarization and (b) the relative tunnel-coupling strength  respectively. The pumping parameters are $\epsilon$ and $\pol$. 
	}
	\label{fig_spin}
\end{figure}

The dependence of the pumped charge and spin on the average polarization $\bar P$ and the ratio of the tunnel coupling to the ferromagnet and the total coupling, $\bar\Gamma_\text{F}/\bar \Gamma$, is shown in Figs.~\ref{fig_charge}~and~\ref{fig_spin}, respectively.
The electrical charges and spins flow in different directions, thus the particle and spin currents flow in the same direction. 

We find that both the pumped charge and spin vanish for $\bar\Gamma_\text{F}/\bar \Gamma$ going to zero, quadratically the former and linearly the latter. 
For  $\bar P$ going to zero the pumped charge goes linearly to zero, while the pumped spin remains finite. In this limit, the pumping scheme  generates a pure dc spin current, i.e. a finite spin current with no associated charge current.  
The spin-pumping efficiency, defined as the pumped spin per pumped charge, can be obtained immediately from Eqs. (\ref{charge-spin}) and it reads 
%
%
\begin{equation}
\label{ratio N}
	R=-2e\dfrac{S_{\epsilon,P}}{Q_{\epsilon,P}}= 
	\dfrac{1}{2}\left( \dfrac{\bar{\g}}{\bar{\pol} \bar{\g}_{\rm{F}}}+\dfrac{\bar{\pol} \bar{\g}_{\rm{F}}}{\bar{\g}}\right) \, .
\end{equation}
It is worth noticing that it becomes arbitrarily large for $P\bar\Gamma_\text{F}/\bar \Gamma \rightarrow 0$. In the limit   $P\bar\Gamma_\text{F}/\bar \Gamma \rightarrow1$ the number of pumped charges equals the one of the pumped spins.

\subsubsection{Contribution from pumping with $\epsilon$ and $\Gamma_{\rm{F}}$}

\begin{figure}
	\includegraphics[width=0.45\textwidth,angle=0,clip]{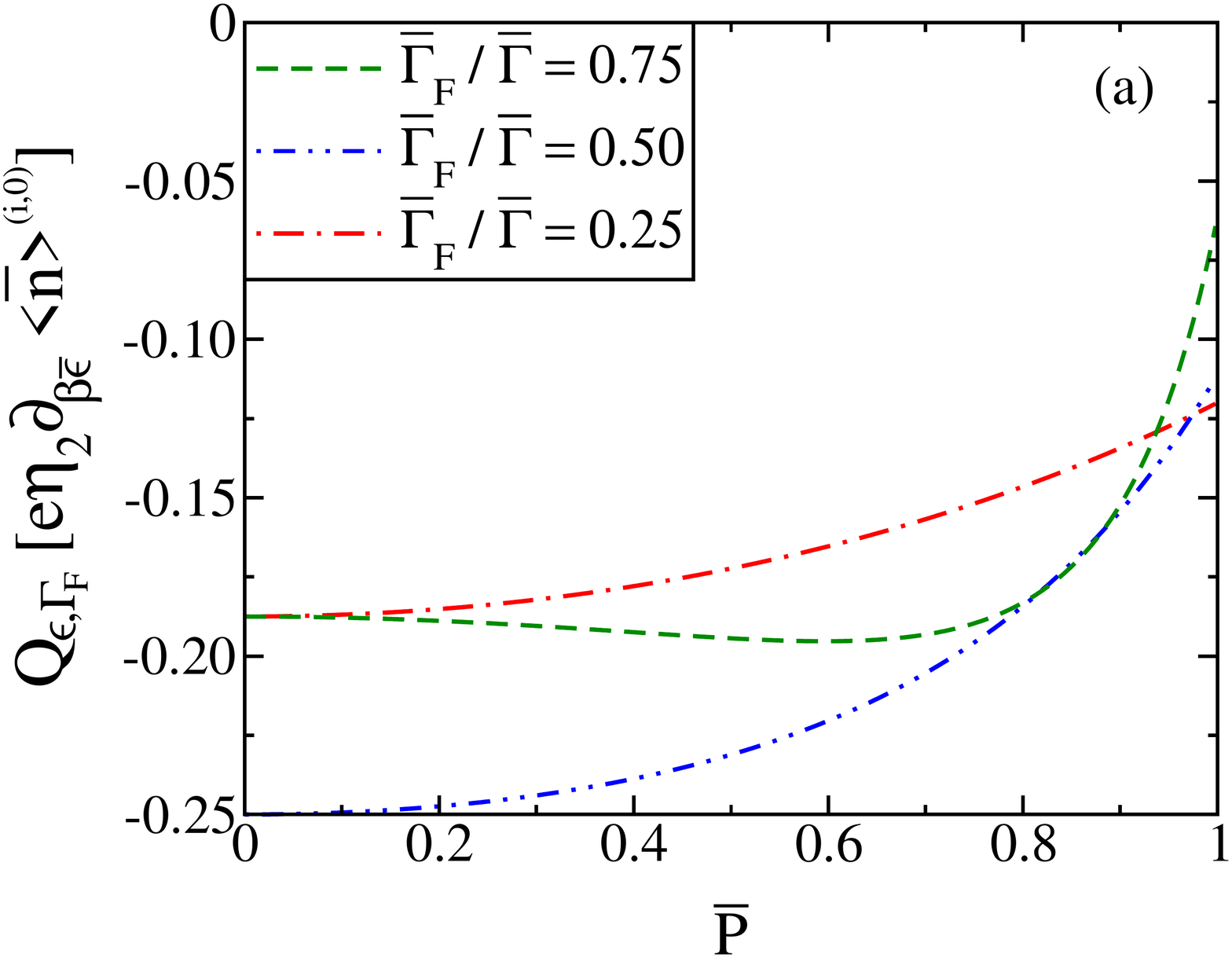} \\
	\includegraphics[width=0.45\textwidth,angle=0,clip]{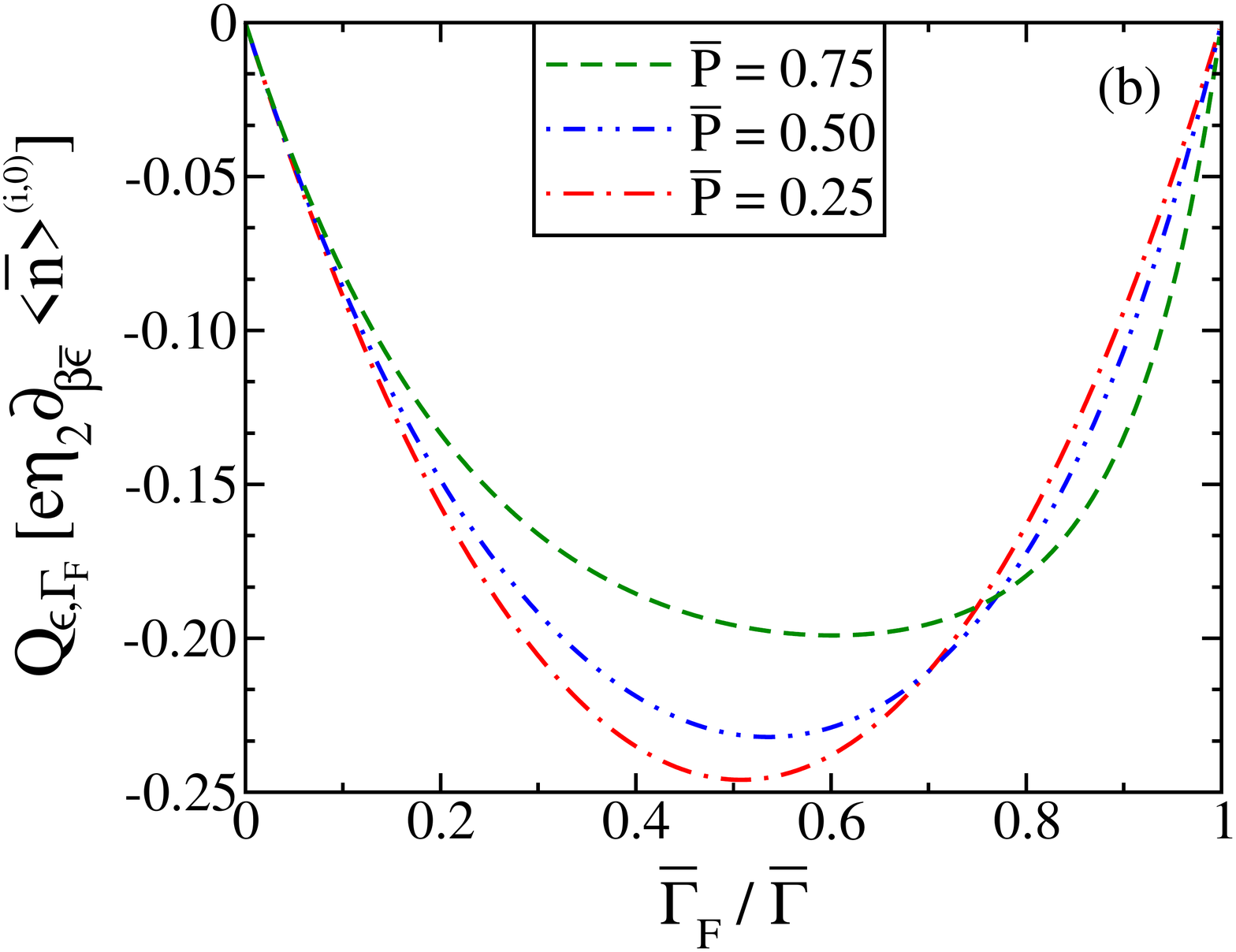}\\
	\caption{
	(Color online)  
	Pumped charge in units of $e \eta_2 \partial_{\beta \bar{\epsilon}}   \left\langle \bar{n}\right\rangle ^{(\text{i},0)}  $ 	as a function of (a) the time average of the  lead polarization and (b) the relative tunnel-coupling strength  respectively. The pumping parameters are $\epsilon$ and $\gf$. 	
	}
	\label{fig_charge_gf}
\end{figure}

\begin{figure}
	\includegraphics[width=0.45\textwidth,angle=0,clip]{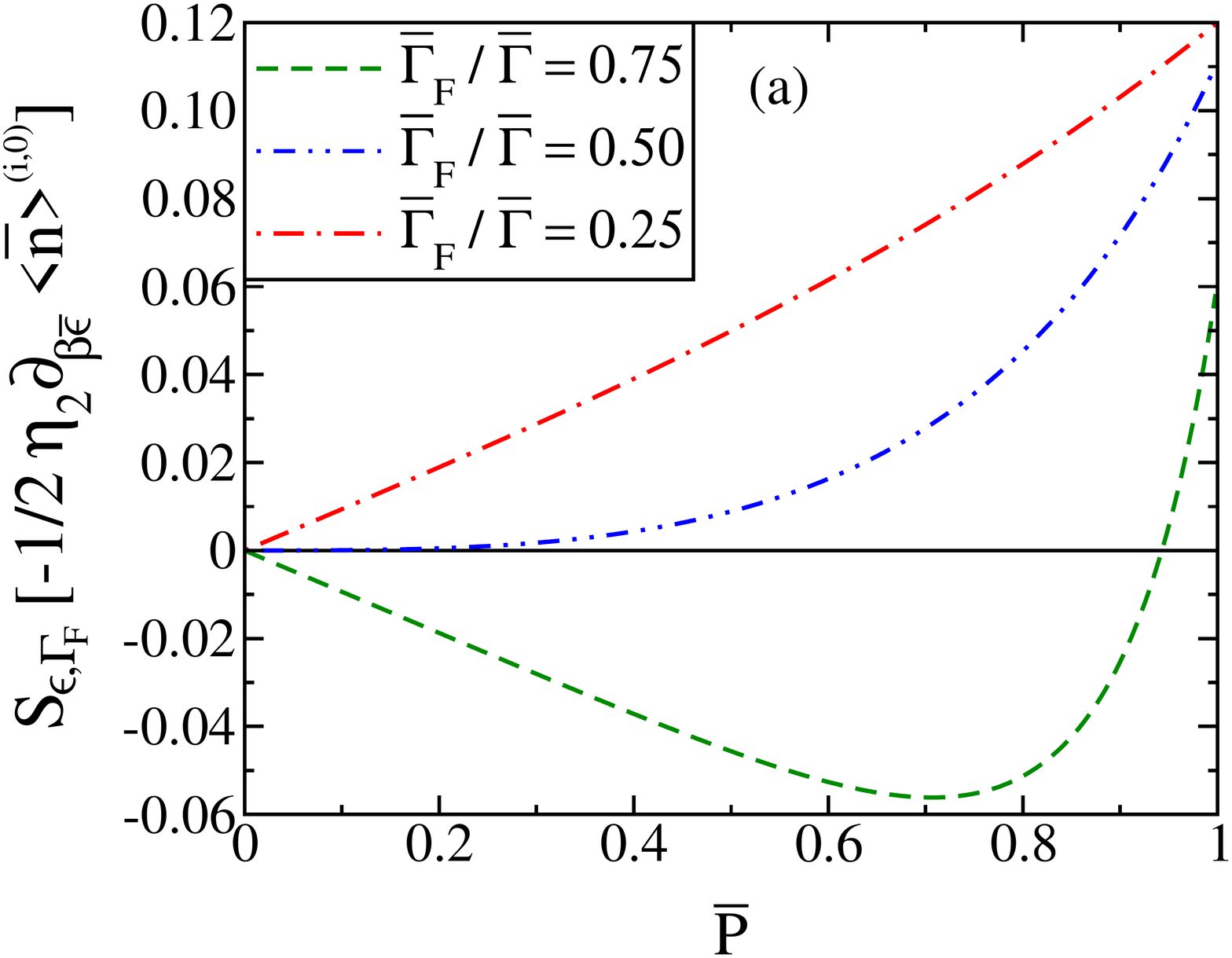} \\
	\includegraphics[width=0.45\textwidth,angle=0,clip]{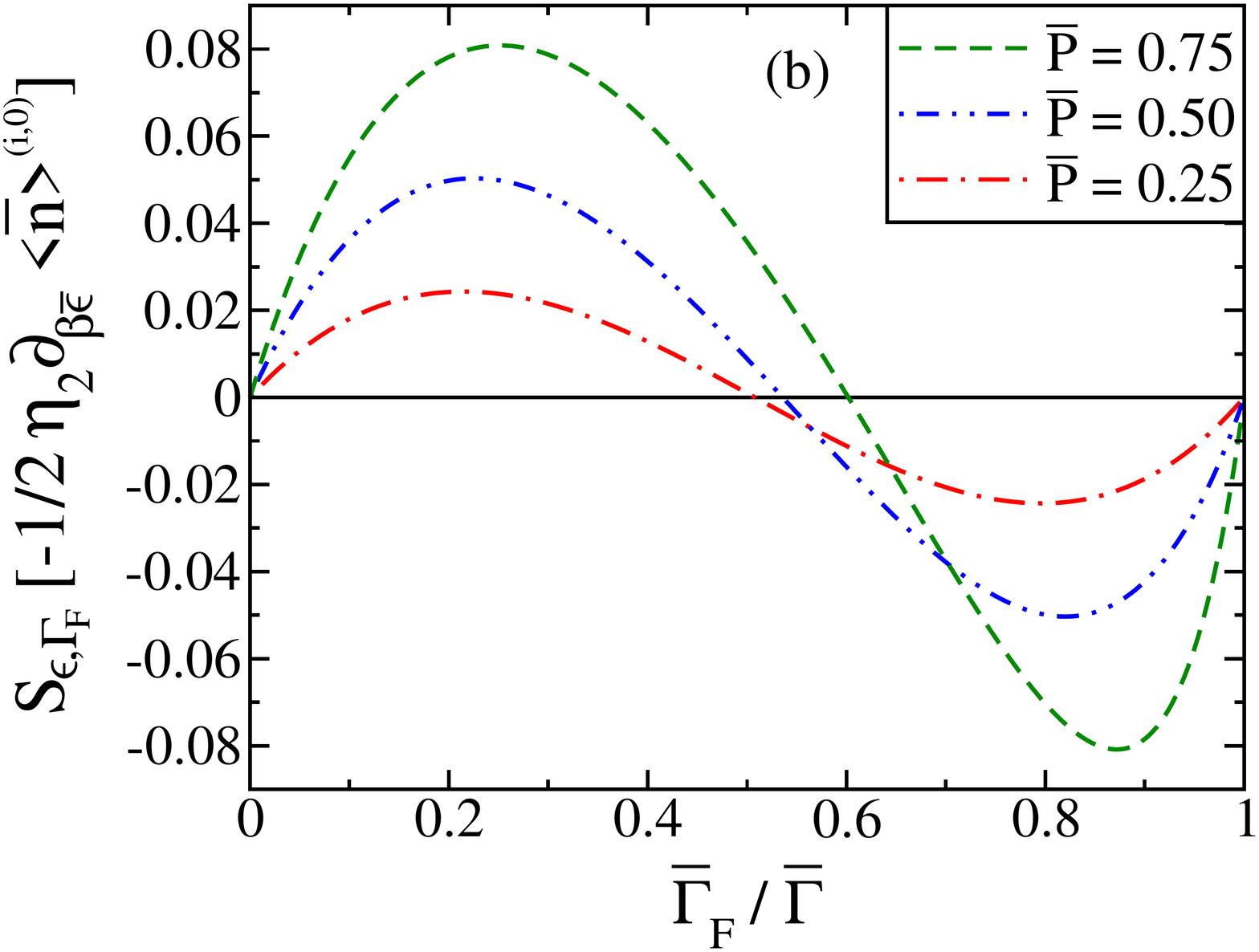}\\
	\caption{
	(Color online) 
	Pumped spin in units of $-\frac{1}{2} \eta_2 \partial_{\beta \bar{\epsilon}}   \left\langle \bar{n}\right\rangle ^{(\text{i},0)}  $ 	as a function of (a) the time average of the  lead polarization and (b) the relative tunnel-coupling strength  respectively. The pumping parameters are $\epsilon$ and $\gf$. 
	}
	\label{fig_spin_gf}
\end{figure}

Next, we consider the contribution to the pumped charge and spin that originates from pumping with $ \epsilon$ and $\gf $. This contribution coincides with  the results obtained for pumping by changing the properties of the scattering region  ($ \epsilon$ and $\gf $)  in  a F-dot-N structure, investigated in Ref.~\onlinecite{splettstoesser_adiabatic_2008}.
In the limit of weak pumping, we find
%
%
\begin{subequations}
\begin{align}
\label{}
	Q_{\epsilon, \gf} &=  e    \eta_2 
	  \dfrac{ \bar{\pol}^2  \, \frac{\bar{\Gamma}_{\rm{F}}^{2}}{\bar{\Gamma}^2}\, \frac{\bar{\Gamma}_{\rm{N}}}{\bar{\Gamma}} \left( 2 - \frac{\bar{\Gamma}_{\rm{F}}}{\bar{\Gamma}} \right)   - \frac{\bar{\Gamma}_{\rm{F}}}{\bar{\Gamma}}\, \frac{\bar{\Gamma}_{\rm{N}}}{\bar{\Gamma}}  }	{\left( 1- \bar{\pol}^2 \, \frac{\bar{\Gamma}_{\rm{F}}^2}{\bar{\Gamma}^2} \right)^{2}} \; \dfrac{d}{d\beta	\bar{\epsilon}}   \left\langle \bar{n}\right\rangle ^{(\text{i},0)}  \\
\label{}
	S_{\epsilon, \gf} &=- \frac{1}{2}   \eta_2 
	  \dfrac{\bar{\pol} \frac{\bar{\Gamma}_{\rm{F}}}{\bar{\Gamma}}\, \frac{\bar{\Gamma}_{\rm{N}}}{\bar{\Gamma}}  \left( 1+ \bar{\pol}^2 \, \frac{\bar{\Gamma}_{\rm{F}}^2}{\bar{\Gamma}^2} - 2 \frac{\bar{\Gamma}_{\rm{F}}}{\bar{\Gamma}} \right)}	{\left( 1- \bar{\pol}^2 \, \frac{\bar{\Gamma}_{\rm{F}}^2}{\bar{\Gamma}^2} \right)^{2}} \; \dfrac{d}{d\beta	\bar{\epsilon}}   \left\langle \bar{n}\right\rangle ^{(\text{i},0)}  ,
\end{align}
\end{subequations}
where $\eta_2 =\int _0^{\mathcal{T}} dt \; \delta\beta\epsilon\; \delta \dot{\Gamma}_\text{F}/\bar{\Gamma}_\text{F} $ is the area of the pumping cycle in parameter space, normalized by $\bar{\Gamma}_\text{F}$.
The dependence of the pumped charge and spin as a function of $\bar P$ and $\bar\Gamma_\text{F}/\bar \Gamma$ is shown in Figs.~(\ref{fig_charge_gf}) and~(\ref{fig_spin_gf}).
As already remarked in Ref.~\onlinecite{splettstoesser_adiabatic_2008}, the pumped spin changes signs as a function of 
$\bar\Gamma_\text{F}/\bar \Gamma$, while the pumped charge 
does not. 
 %
 %
%
%
%
%

%
%
\begin{figure}
	\includegraphics[width=0.475\textwidth,angle=0,clip]{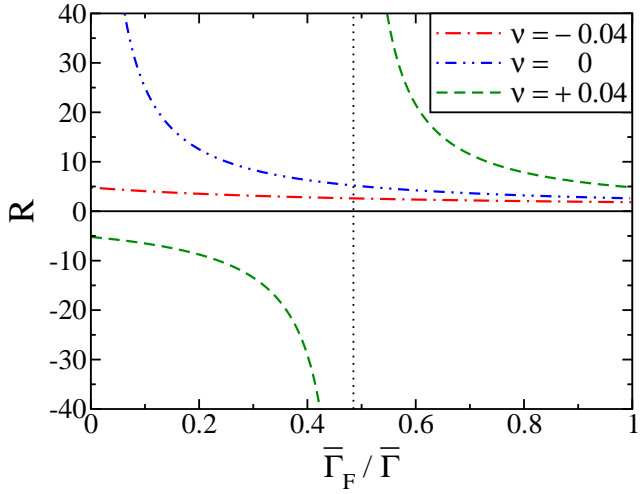} \\
	\caption{Total efficiency of the spin pump as a function of the time averaged ratio $\bar{\g}_{\rm{F}}/\bar{\g}$ plotted for different values of $\nu$ when the polarization is chosen to be $P=0.2$\;.}
	\label{fig_sum_ratio}
\end{figure}

\subsubsection{Total pumping efficiency and pure spin current}

After having discussed separately the two contributions due to  pumping with $\{ \epsilon, P\} $ and $\{\epsilon, \gf\}$, we now turn to address the sum of the two.
In Fig.~(\ref{fig_sum_ratio}), we show the total spin efficiency  
\begin{equation}
 	R=-2e\dfrac{S_{\epsilon,P} + S_{\epsilon,\gf}}{Q_{\epsilon,P} + Q_{\epsilon,\gf}}  \, ,
\end{equation}
as a function of $\bar{\g}_{\rm{F}}/\bar{\g}$ for different values of $\nu$ and a polarization $P=0.2 $.
For $\nu=0$ 
(realized for flat bands around the Fermi energy) we get only the contribution due to pumping with $\epsilon$ and $P$ which diverges for 
$\bar{\g}_{\rm{F}}/\bar{\g} \rightarrow 0$.
A divergence of the spin efficiency 
indicates a pure spin current without a charge current.
However, for $\nu=0$
the pure spin current is only asymptotically reached for $\bar{\g}_{\rm{F}}/\bar{\g} \rightarrow 0$ since in this limit the amplitude of the spin current vanishes.
A negative value of $\nu$ removes the divergency (no pure spin current).
The most interesting case is realized for positive values of $\nu$. 
In this case, the divergence of the spin efficiency is shifted to finite values of $\bar{\g}_{\rm{F}}/\bar{\g}$, which correspond to a pure spin current of finite amplitude.
The sign change of $R$ at this point reflects a sign change in the total pumped charge current.

\subsection{Pumping scheme B: rotating lead magnetizatiom}

In pumping scheme B, the magnitude $M$ of the ferromagnet's magnetization $\vec{M}(t)=M {\hat{e}}_p(t)$ remains fixed but its direction ${\hat{e}}_p(t)= \left( \sin\theta \, \cos\varphi(t), \sin\theta \, \sin\varphi(t), \cos\theta \right)^\text{T}$ rotates about the $z$-axis.
The latter is used as the (time-independent) quantization axis for the dot and normal lead electron spins $\sigma=\up,\down$, whereas the direction of the majority and minority spins $\alpha=\pm$ of the ferromagnet changes in time.  
The Stoner splitting $\Delta E = E_{k-} - E_{k+}$, on the other hand, remains constant in time. 
To experimentally induce such a rotation in the ferromagnet, one may make use of a ferromagnetic resonance.
 
As mentioned above, adiabatic pumping requires the time-variation of two system parameters with a relative phase.
In pumping scheme B, the two parameters are the x- and y-component of the polarization, $P_x(t)=P \sin \theta \cos \varphi (t)$ and $P_y(t)=P \sin \theta \sin \varphi (t)$. 
In contrast to pumping scheme A, we do not need to vary the dot-level position in order to achieve pumping. 
 
In pumping scheme B, the azimuthal angle $\varphi(t)$ of the ferromagnet's magnetization direction is the only time-dependent parameter.
As a consequence, the instantaneous average dot occupation $\langle n \rangle^{(i,0)}$ is constant in time.
From Eqs.~(\ref{ccur_a_0}) and (\ref{scur_a_0}) we deduce that both the charge and the spin current vanish to lowest order in the tunnel coupling, $I_{\text{N}}^{(\text{a},0)}(t) = 0$ and $\vec{J}_{\text{N}}^{(\text{a},0)}(t) = \mathbf{0}$.
It is, therefore, necessary to include the next-order contribution in the perturbation expansion in the tunnel-coupling strength.

An explicit calculation, presented in Appendix~\ref{appendix_first_order_B}, yields a vanishing pumped charge current, $I_\text{N}^{(\text{a},1)} = 0$, but a finite pumped spin current, which can be nicely written in a compact analytical form, 
\begin{widetext}
\begin{align}
\label{scur_a_1}
\mathbf{J}_\text{N}^{(\text{a},1)}
&= \,- \,\frac{\Gamma_{\,\text{F}}\Gamma_{\,\text{N}}}{4\Gamma^2}  \;  \frac{ \partial_\epsilon   \langle n \rangle^{(\text{i},0)}   }{\tau^Q_\text{rel} }  \;  
\left\{
\left( 1\;+\;  \frac{\Gamma_{\text{N}}}{\Gamma_{\text{F}}}
\; \frac{\left(B \; \tau^S_\text{rel}\right)^2}{1 +    \left(B \; \tau^S_\text{rel}\right)^2 } \right)  \;   \ep  \times \dtep  
\;-\;  \frac{\Gamma_{\text{N}}}{\Gamma_{\text{F}}}
\; \frac{ B \; \tau^S_\text{rel} }{1+    \left(B \; \tau^S_\text{rel}\right)^2 }   \;   \dtep  \right\}   \,.
\end{align}
\end{widetext}
Here, $\tau^Q_\text{rel}$ and $\tau^S_\text{rel}$ are the charge and spin relaxation times, respectively, and $B$ describes an interaction-induced exchange field that is a consequence of the spin-dependent tunnel coupling of the dot level to the ferromagnet.\cite{koenig_interaction_2003,martinek_kondo_qd_2003,braun_theory_2004}
The explicit expressions are given by
\begin{align}
%
%
\frac{1}{\tau^Q_\text{rel}} 
&= \Gamma \left[f^+(\epsilon)+ f^-(\epsilon+U)  \right] 
\\
%
%
\frac{1}{\tau^S_\text{rel}} 
&= \Gamma \left[f^-(\epsilon)+ f^+(\epsilon+U)  \right] 
\\
%
B &=  \frac{\Gamma_{\text{F}} \, P }{\pi} \; \pint d\omega  \left[ \frac{f^- (\omega)}{ \omega - \epsilon }  \;+\;   \frac{f^+ (\omega) }{ \omega - \epsilon - U } \right] \, ,
\end{align}
where $\g=\gn+\gf$ is the total tunnel-coupling strength, and $ \pint \,\, d\omega$ denotes Cauchy's principal value.

To get the pumped spin per cycle we need to integrate over one pumping cycle.
From the symmetry of the problem it is clear that the spin components along the $x$- and $y$-directions average out, and only the $z$-component survives. 
Since $\dtep$ does not have any $z$-component, it is only the term with $\ep \times \dtep$ that contributes to the finite pumped spin per cycle $S_\varphi$.
We use $(\ep \times \dtep) \cdot \hat e_z = \dot\varphi(t) \sin^2 \theta $ and assume a constant angular velocity, $\Omega= \dot{\varphi}(t)$, to get 
\begin{align}
S_\varphi = \frac{1}{4} \Omega \sin^2\theta \,G_0 \left( 1 +  \frac{\gn}{\gf}
 \frac{\left(B \tau^S_\text{rel}\right)^2}{1 +  \left(B \tau^S_\text{rel}\right)^2 } \right)   
\label{eq_spin_current_a1}
\end{align}
expressed with the help of the dimensionless linear conductance (in units of $e^2/h$)
\begin{equation}
	G_0= -2\pi \;\frac{\gn \; \gf}{\g^2} \;  \frac{\partial_\epsilon    \langle n \rangle^{(\text{i},0)}}{\tau^Q_\text{rel}}  \;   
\end{equation}
of the F-dot-N structure for vanishing polarization. 

The exchange field $B$ enters the expression for the pumped spin, Eq.~(\ref{eq_spin_current_a1}), in combination with the spin relaxation time $\tau^S_\text{rel}$.
This is quite natural since the latter defines the time during which the exchange field acts on the quantum-dot spin before it relaxes due to the dot electron leaving the dot or another electron entering the dot via tunneling to or from the leads, respectively.
It is remarkable that the presence of the exchange field {\it enhances} the pumped spin.
To understand this, we notice that the exchange field affects the spin dynamics in two ways (for a mathematical support of the following physical argument, see also Eq.~(\ref{kin-eq_sa0_1})).
It is clear that the exchange field should induce a precession of an accumulated quantum-dot spin about $\hat e_p$ (last term in Eq.~(\ref{kin-eq_sa0_1})).
If this were the only effect, the pumped spin into the normal lead would read $S_\varphi = \frac{1}{4} \Omega \sin^2\theta \,G_0 / \left[1 +  \left(B \tau^S_\text{rel}\right)^2 \right]$, i.e., the pumped spin would be {\it reduced} by this precession.
There is, however, another contribution in which the exchange field enters, namely the accumulation term (first term in Eq.~(\ref{kin-eq_sa0_1})).
This indicates that already during the generation of the accumulated spin, the spin dynamics induced by the exchange field plays a role and gives a finite contribution to the adiabatic correction to the spin accumulation.
As a result of our calculation, we find that the combination of the two effects leads to the form in Eq.~(\ref{eq_spin_current_a1}) for the pumped spin which increases with increasing exchange field.

\begin{figure}
	\includegraphics[width=0.45\textwidth,angle=0,clip]{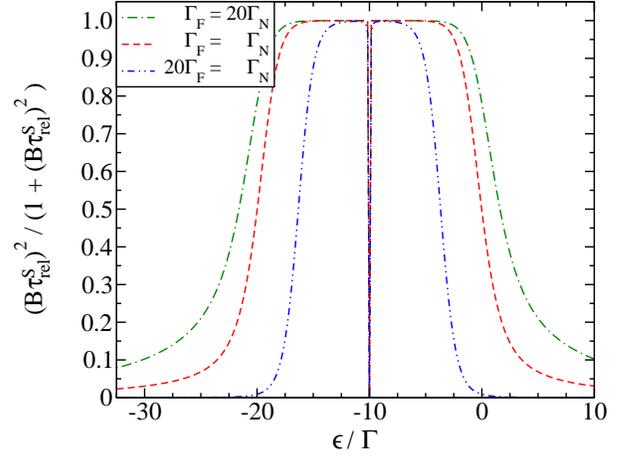} 
	\caption{(Color online) The factor $\left(B \; \tau^S_\text{rel}\right)^2 / \left[ 1 +    \left(B \; \tau^S_\text{rel}\right)^2 \right] $ plotted as a function of the dot level position $\epsilon$ for different choices of the tunnel couplings, fixed Coulomb energy $U=20\g$, and fully polarized ferromagnetic lead ($P=1$).}
	\label{fig_factor}
\end{figure}

Due to the prefactor $\gn/\gf$ in front of the term $\left(B \; \tau^S_\text{rel}\right)^2 / \left[ 1 +    \left(B \; \tau^S_\text{rel}\right)^2 \right] $,  the exchange field becomes more and more important when increasing the tunnel coupling to the normal lead.
The latter term is plotted as a function of the dot level $\epsilon$ in Fig.~\ref{fig_factor} for a constant value of the  Coulomb on-site energy $U=20\g$.
%
%
It goes to zero for large values of $|\epsilon |$ and has two maxima symmetrically positioned around its minimum at $\epsilon = - \frac{U}{2}$. With increasing Coulomb interaction $U$, the maxima reach the value one and the width of the peaks increases, which leads for large values of the  Coulomb interaction, as it is the case chosen for Fig.~\ref{fig_factor}, to a plateau with a slit. 

\begin{figure}
	\includegraphics[width=0.45\textwidth,angle=0,clip]{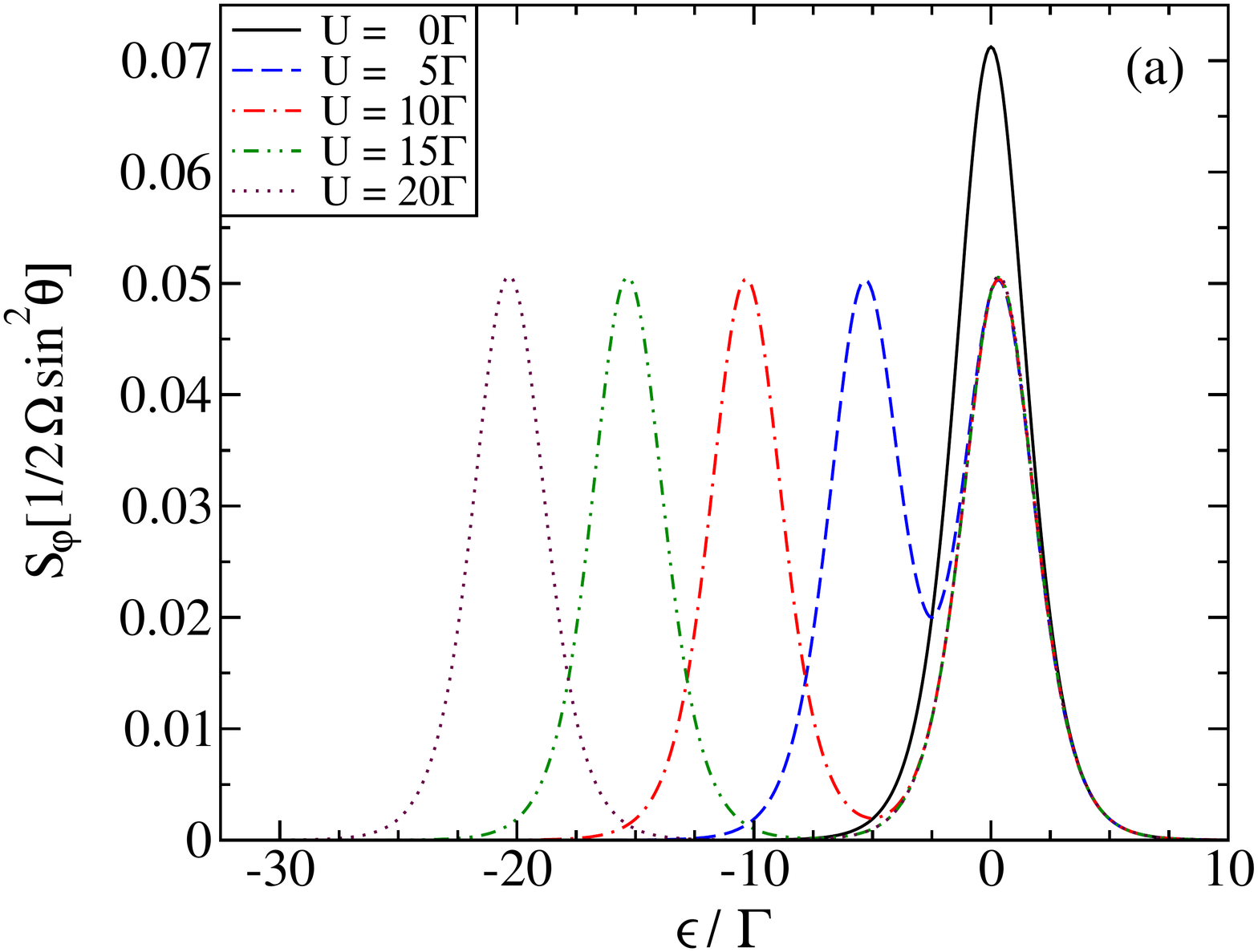} 
	\includegraphics[width=0.45\textwidth,angle=0,clip]{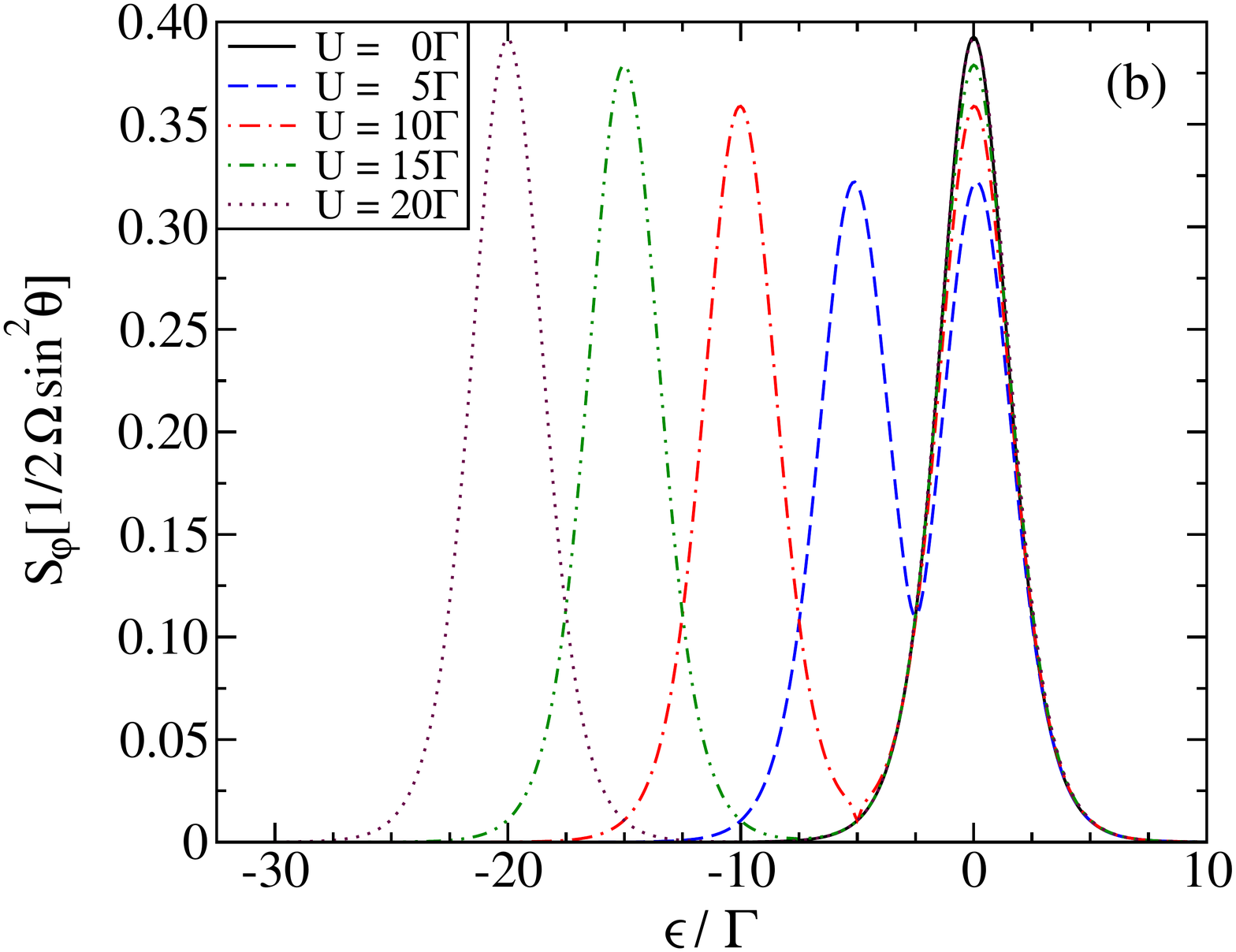} 
	\includegraphics[width=0.45\textwidth,angle=0,clip]{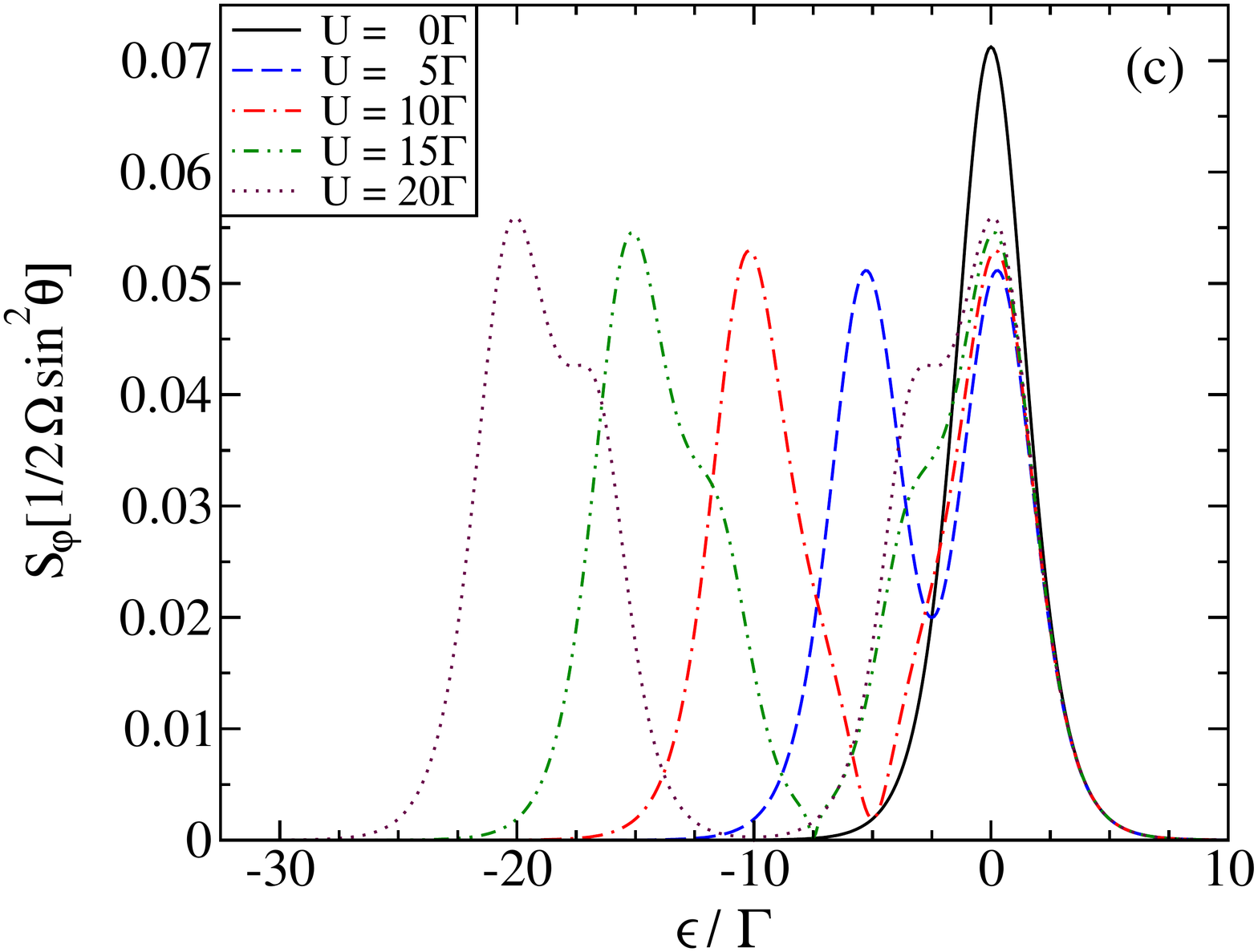} 
	\caption{(Color online) Pumped spin in units of $\frac{1}{2}  \Omega \sin^2\theta$ 	as a function of the level position $\epsilon$ of the dot for different values of the Coloumb 		interaction $U$. The temperature is chosen to be $k_\text{B}T= \Gamma$ and the 			ferromagnetic lead is assumed to be fully polarized ($\pol=1$).	\\
	(a) The coupling to the ferromagnetic lead is strong \mbox{$(\gf=20\gn)$}. \\ (b) The coupling to the leads is chosen symmetric $(\gf=\gn)$. \\ (c)  The coupling to the ferromagnet is weak 	$(20\gf=\gn)$.  }
	\label{fig_spin_current}
\end{figure}

In Fig.~\ref{fig_spin_current}, we illustrate how the effect of the exchange field depend on the tunnel couplings.
We begin by choosing a weak tunnel coupling to the normal lead. 

Then, the second term of the pumped spin including the exchange field is small compared to the first term and does not play a role. The pumped spin in this situation is plotted as a function of the dot level $\epsilon$ in panel (a) of Fig.~\ref{fig_spin_current} for different values of the charging energy $U$. In the absence of  Coulomb interaction on the dot, there is one peak positioned at $\epsilon=0$. In the presence of the Coulomb interaction on the dot, a second resonance appears at $\epsilon=-U$ while the amplitude of the maxima is decreased and stays constant for Coulomb interactions unequal to zero.

Next, we choose the tunnel couplings symmetrically, $\gf = \gn$.
A plot of the spin in this case is shown in panel (b) of Fig.~\ref{fig_spin_current}. The symmetric choice of the tunnel couplings makes the amplitude of the pumped spin maximal. 
One can see that the resonances are still at the same positions and that only the amplitude of the maxima decreases at first but increases with increasing Coulomb interaction. This is due to the enhancement of the factor with increasing $U$. The shape of the factor leads as well to a dip at the minimum for $U=10\g$.

The pumped spin for asymmetric tunnel couplings with a strong coupling to the normal lead is plotted in panel (c) of Fig.~\ref{fig_spin_current}. Compared to the symmetric case, the overall amplitude has been reduced and it increases more slowly with increasing Coulomb interaction. The positions of the main resonances still coincide. For this choice of the tunnel couplings the exchange field plays a crucial role. 
The value of the pumped spin is increased for gate voltages between the resonances. This gives rise to side peaks, one below $\epsilon=0$ and one above $\epsilon=-U$. These peaks overlap with the corresponding main peaks. They are visible as individual peaks only for large enough values of the Coulomb interaction $U$. 

We remark that the polarization of the ferromagnet enters the expression of the pumped spin current~(\ref{eq_spin_current_a1}) only via the magnitude of the exchange field $B$. 
For vanishing Stoner splitting $\Delta E$ and, thus, the polarization going to zero, Eq.~(\ref{eq_spin_current_a1}) would still give a finite result. However, in this case the assumption, discussed in Section \ref{formalism}  that the spin-relaxation 
time in the ferromagnet  needs to be longer the the time it takes to complete a transitions between minority and majority states is not verified and Eq.~(\ref{eq_spin_current_a1}) cannot be applied any longer. When this happens the spin flip process in the ferromagnet,  described by tunneling lines contacted with a double-cross vertex,  are cut by the spin relaxation time in the ferromagnet.

\section{Conclusions}
\label{sec_conclusion}

We studied adiabatic charge and spin pumping through a single-level quantum dot weakly tunnel-coupled to a normal and a ferromagnetic lead with time-dependent polarization.
To this end, we extended a real-time diagrammatic approach to account for a time variation of the ferromagnet's properties. 
We investigated two different pumping schemes. In the first one, the amplitude of the ferromagnet's polarization is changed in time.
To establish pumping, we chose the dot's level position as a second pumping parameter.
A pure spin current without any charge current is only possible for special choices of the system parameters.  
The second pumping scheme relies on the rotation of the magnetization direction of the ferromagnet. 
In this case, the pumped charge current always vanishes, i.e., 
a pure spin current is generated
without fine tuning of the system parameters.
 
We acknowledge financial support from DFG via SPP 1285 and from EU under Grant No. 238345 (GEOMDISS).

\begin{appendix}

\section{Kinetic equations in case of rotational spin symmetry}
\label{appendix_general_trafo}

We consider the equations of the from $\frac{d}{dt}\;\mathbf{p}  \;=\; \mathbf{W}  \;\mathbf{p}$, where the vector $\mathbf{p}$ of matrix elements of the reduced density matrix is given by $\mathbf{p}=(p_0,p_\uparrow,p_\downarrow,p_d,p^\downarrow_\uparrow,p^\uparrow_\downarrow)^\text{T}$.
Within this appendix, the shortcut notation $\mathbf{W}  \;\mathbf{p} $ could either represent the time convolution
$\int\limits_{-\infty}^{t}dt'\;\mathbf{W} \left( t,t'\right)\;\mathbf{p}(t')$ or the product $\mathbf{W}_t^{(i)} \;\mathbf{p}_t^{(i)} $ for the kinetic equation in the instantaneous limit.

The basis change from $\mathbf{p}$ to $\mathbf{P}$ and $\mathbf{S}$ is accomplished by
the transformation
\begin{equation}
\label{transformation}
	\left( \begin{array}{c} \mathbf{P} \\ \mathbf{S} \end{array} \right)
	=
	\left(
	\begin{array}{cccccc} 
	1 & 0 & 0 & 0 & 0 & 0 \\ 
	0 & 1 & 1 & 0 & 0 & 0 \\ 
	0 & 0 & 0 & 1 & 0 & 0 \\ 
	0 & 0 & 0 & 0 & 1/2 & 1/2 \\ 
	0 & 0 & 0 & 0 & -i/2 & i/2\\ 
	0 & 1/2 & -1/2 & 0 & 0 & 0
	\end{array} \right)
	\;
	\mathbf{p}
	\; .
\end{equation}
In case of rotational spin symmetry about the axis $\hat e_p$, it is convenient for the following derivation to quantize the spin along this symmetry axis.
In that basis the kernel $\mathbf{W}$ reads:
\renewcommand{\arraystretch}{1.75}
\begin{align*}
%
%
%
%
%
%
\mathbf{W} \; & =    
 \left(
\begin{array}{cccccc}
%
 W_{00} &W_{0\up} &W_{0\down} &W_{0d} & 0 & 0 \\	
%
W_{\up 0} & W_{\up \up} & W_{\up \down} & W_{\up d} & 0 & 0 \\	
%
W_{\down 0} & W_{\down \up} & W_{\down \down} & W_{\down d} & 0 & 0 \\ 
%
W_{d0} & W_{d \up} & W_{d \down} & W_{d d} & 0 & 0 \\ 
%
0 & 0 & 0 & 0 & W^{\down \down}_{\, \up \up} & 0\\ 
%
0 & 0 & 0 & 0 & 0 &  W^{\up \up}_{\, \down \down} 
\end{array}
\right) \, ,
\end{align*}
where the zeros for the off diagonal matrix elements in the fifth and sixth column and row are a consequence of the spin symmetry.

As a result, the kinetic equations for $\mathbf{P}$ and $\mathbf{S}$ read 
\begin{align*}
\frac{d}{dt}\;\mathbf{P}
&= \mathbf{W}_p \;  \mathbf{P} + \mathbf{v}_p \;(\mathbf{S}  \cdot \ep) \,,
\\
\frac{d}{dt}\;\mathbf{S}
&= (\mathbf{v}_\text{acc} \cdot \mathbf{P}) \ep - \frac{ \mathbf{S}^\| }{\tau_S^\|   }  
- \frac{ \mathbf{S}^\perp }{\tau_S^\perp }
  +  B \, \mathbf{S} \times \ep  \,,
\end{align*}
where we split the spin vector $\mathbf{S} = \mathbf{S}^\| + \mathbf{S}^\perp$ into a parallel,
$ \mathbf{S}^\| = \left( \mathbf{S}\cdot \ep \right) \ep$, and perpendicular part, $ \mathbf{S}^\perp = \mathbf{S} - \mathbf{S}^\| $, and we made use of the abbreviations
\begin{align*}
%
%
%
\mathbf{W}_p \; & =    
 \left(
	\begin{array}{ccc}
  	W_{00} & \frac{1}{2} \sum_\sigma W_{0 \sigma} & W_{0d} \\	
 	\sum_\sigma W_{\sigma 0} & \frac{1}{2} \sum_{\sigma\sigma'} W_{\sigma\sigma'} & \sum_\sigma W_{\sigma d} \\	
 	W_{d0} & \frac{1}{2} \sum_\sigma W_{d\sigma} & W_{dd}  
\end{array}
\right) \\ 
\\
%
%
%
\mathbf{v}_p \; & =    
 \left(
\begin{array}{c}
W_{0\up} - W_{0\down} \\
\sum_\sigma \left( W_{\sigma\up} - W_{\sigma\down} \right) \\
W_{d\up} - W_{d\down} 
\end{array}
\right) \,,
\\
%
%
\mathbf{v}_\text{acc} \; & =    
\left(
\begin{array}{c}
\frac{1}{2} \left(W_{\up0} - W_{\down0}\right) \\
\frac{1}{4} \sum_\sigma \left( W_{\up\sigma} - W_{\down\sigma}  \right) \\
\frac{1}{2} \left(W_{\up d} - W_{\down d}\right)
\end{array}
\right)  \,
\\
%
%
 \frac{1}{\tau_S^\|   }  
&= -\frac{1}{2} \left( W_{\up\up} - W_{\down\up} + W_{\down\down} - W_{\up\down} \right) \\
 \frac{1}{\tau_S^\perp   }  
 &= -\frac{1}{2} \left(  W^{\up\up}_{\, \down\down} + W^{\down\down}_{\, \up\up} \right)  =  -\Re W^{\up\up}_{\, \down\down} \\
%
%
%
B 
&= -\frac{i}{2} \left(  W^{\up\up}_{\, \down\down} - W^{\down\down}_{\, \up\up} \right)  =  \Im  W^{\up\up}_{\, \down\down} \,.
\end{align*}

The right hand side of kinetic equation for the spin is split into three parts.
The first one is independent of $\mathbf{S}$ and describes spin accumulation.
The second one models the relaxation of the parallel and perpendicular components of the accumulated spin.
And finally, the third term gives rise to a coherent rotation of the spin.

We observe that the spin symmetry has several consequences: only the spin component parallel to the symmetry axis $\ep$ enters the kinetic equation for $\mathbf{P}$, the spin accumulation is along $\ep$, the spin relaxation 
is rotationally symmetric about $\ep$, and the spin rotation is about the symmetry axis $\ep$. 
 
\section{Examples of diagrams}
\label{appendix_example_diagrams}

\begin{figure}
	\includegraphics[width=0.35\textwidth,angle=0]{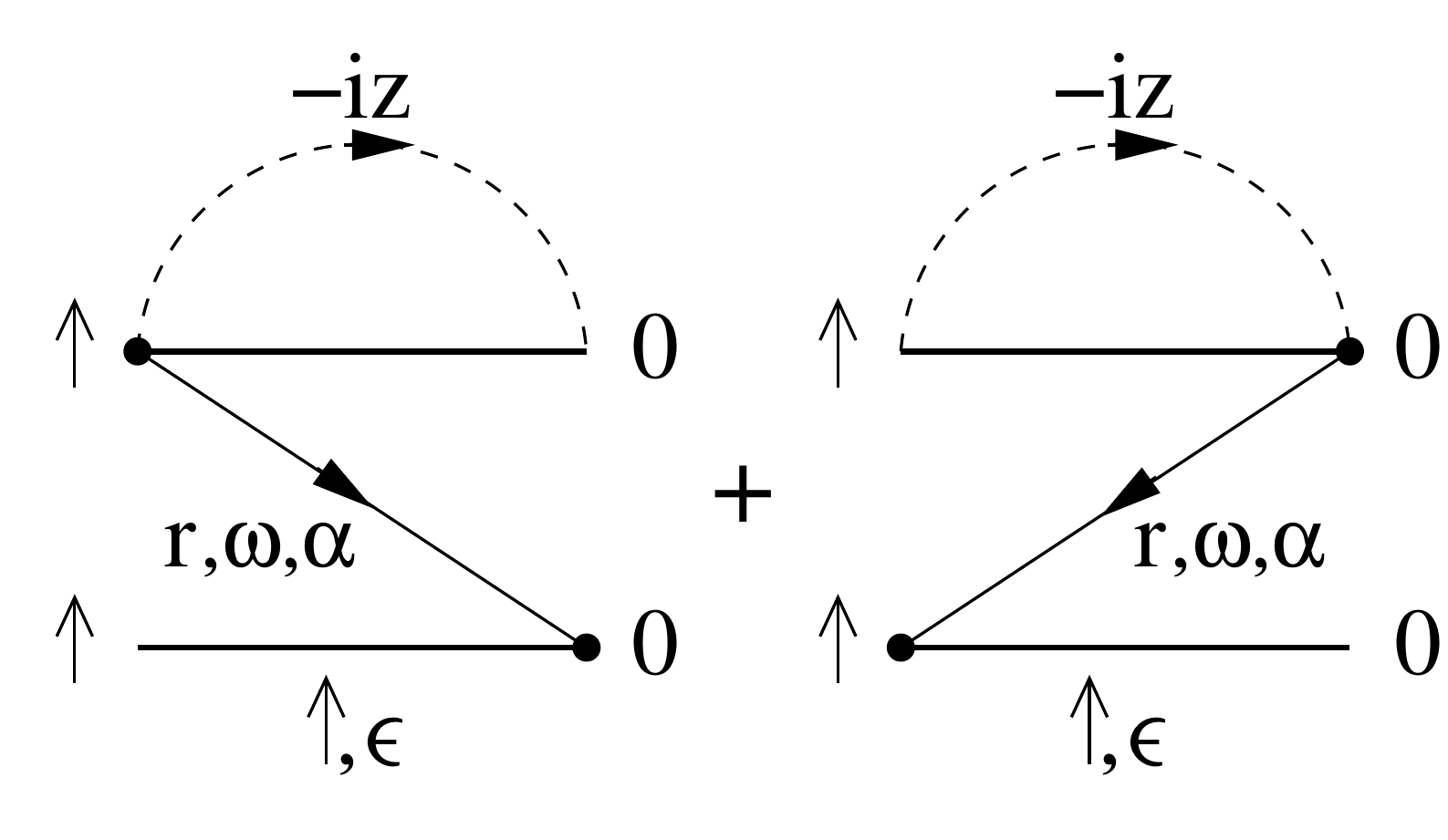}
	\caption{Example of the diagrams which are needed to calculate the matrix element $(\mathbf{W}_t^{(i,	1)})_{0\up}$ belonging to the instantaneous kernel in first order in the tunnel-coupling strength~$\g$.}
	\label{fig_inst_diagram_ex}
\end{figure}

The diagrams contributing to the matrix element $(\mathbf{W}_t^{(i,1)})_{0\up}$ in instantaneous and first order in $\g$ are shown in Fig.~(\ref{fig_inst_diagram_ex}).
Applying the diagrammatic rules, we obtain 
\begin{align*}
(\mathbf{W}_t^{(i,1)})_{0\up}
&= -2 \int d\omega \;  \Im \left[ \frac{f^- \left( \omega \right)}{ \epsilon - \omega +iz} \right]_{z=0^+}\;\;
 \;  \\
& \qquad\quad  \cdot
\left\{ \frac{\gn}{2\pi} + \frac{\gf(\omega) }{2\pi}  \left[ 1 + P(\omega)\right]  \cos^{2}\frac{\theta}{2} \right.\\
&\qquad\qquad \quad\quad
\left. + \frac{\gf(\omega) }{2\pi}  \left[ 1 - P(\omega)\right] \sin^{2}\frac{\theta}{2}   \right\} \\
&= f^-(\epsilon) \left[ \g + P \, \gf \cos\theta \right] \,.
\end{align*}
%
%

We now calculate the adiabatic correction to the instantaneous kernel in first order in $\g$ 
for finite $\dot \varphi$ but $\dot \theta = 0$
(this is necessary only for pumping scheme B). 
%
\begin{figure*}
	\includegraphics[width=1.0\textwidth,angle=0,clip]{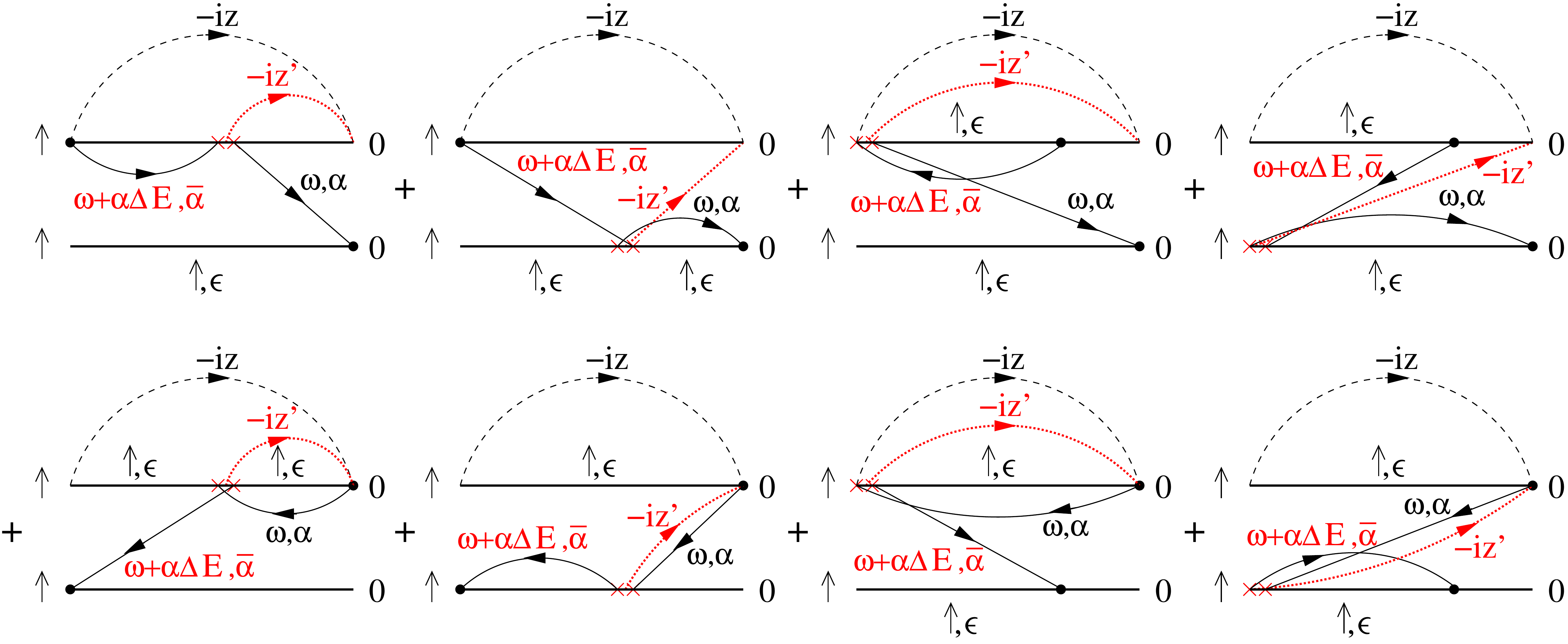}
	\caption{
	(Color online)  Diagrams contributing to the matrix element $( \mathbf{W}_t^{(a,1)} )_{0\up}$ of the adiabatic correction to the kernel in first order of the tunnel-coupling strength $\g$.
	}
	\label{fig_adia_diagrams_ex}
\end{figure*}
Applying the rules~(1.')-(6.') we have to draw for the kernel element $(\mathbf{W}_t^{(a,1)})_{0\up}$ the diagrams shown in Fig.~(\ref{fig_adia_diagrams_ex}) and get 
\begin{widetext}
\begin{align*}
%
%
%
(\mathbf{W}_t^{(a,1)})_{0\up} 
&= \,  \frac{\dot{\varphi}(t)}{2} \; \sin^2\theta   \sum\limits_{\alpha \in \{+1,-1\}}  
 \;  \int d\omega \;\;   \frac{\gf ( \omega)}{2\pi} \;  \left[ 1 + \alpha \, \pol (\omega) \right] \; \alpha \, \Delta E \;\;  \\ 
&\quad \cdot \left\{
\frac{\partial}{\partial z'} \left[   
\Re \left(  \frac{1}{ \epsilon - \omega  - \alpha \, \Delta E  + iz}  \; \frac{1}{\epsilon - \omega + iz'+ iz} \right)  
  \right]_{\substack{z=0^+\\z'=0^+} } 
\cdot
\left[  f^-\left(\omega  \right) \,  f^-\left(\omega + \alpha \, \Delta E \right)  
+ f^+\left(\omega  \right) \,  f^-\left(\omega + \alpha \, \Delta E \right)     \right]     \right. \\ 
&\qquad  \left. 
+\frac{\partial}{\partial z'}  \left[ 
\Re \left(   \frac{1}{\alpha \, \Delta E + iz' +iz}  \; \frac{1}{\epsilon - \omega + iz'+ iz} \right)  
   \right]_{\substack{z=0^+\\z'=0^+} } 
\cdot
\left[  -  f^-\left(\omega  \right)  \, f^+ \left(\omega + \alpha \, \Delta E \right)  
+ f^+\left(\omega  \right)  \,  f^- \left(\omega + \alpha \, \Delta E \right)    \right]     \right\} \\ 
&= \,  \frac{\dot{\varphi}(t)}{2} \; \sin^2\theta  \left[ \gf \; \df (\epsilon) + \frac{2}{\Delta E} \; \pol \, \gf \; f^-(\epsilon)
\right]  \,.
\end{align*}

In the second step, we neglected the energy dependence of $P$ and $\gf$.
It is crucial that before doing this one needs to shift the integration variable $\omega$ such that $\Delta E$ does not appear anymore explicitly in the integral.
For the presented example, this is done by 
\begin{align*}
\int d\omega \;\;   \frac{\gf ( \omega)}{2\pi} \;  \left[ 1 + \alpha \, \pol (\omega) \right] \;   \frac{f^-\left(\omega  \right)}{\epsilon - \omega + i0^+}  
&= \frac{\gf}{2\pi} \;  \left[ 1 + \alpha \, \pol \right] \; \int d\omega \;\;   \frac{f^-\left(\omega  \right)}{\epsilon - \omega + i0^+}  \\ 
\\
\int d\omega \;\;   \frac{\gf ( \omega)}{2\pi} \;  \left[ 1 + \alpha \, \pol (\omega) \right] \;   \frac{f^-\left(\omega  + \alpha \, \Delta E  \right)}{\epsilon - \omega  - \alpha \, \Delta E  + i0^+}  
&=\int d\omega' \;\;   \frac{\gf ( \omega' - \alpha \, \Delta E)}{2\pi} \;  \left[ 1 + \alpha \, \pol (\omega' - \alpha \, \Delta E) \right] \;   \frac{f^-\left(\omega'  \right)}{\epsilon - \omega'  + i0^+}  \\
&=\int d\omega \;\;   \frac{\gf ( \omega)}{2\pi} \;  \left[ 1 - \alpha \, \pol (\omega) \right] \;   \frac{f^-\left(\omega  \right)}{\epsilon - \omega  + i0^+}  \\
&= \frac{\gf}{2\pi} \;  \left[ 1 - \alpha \, \pol \right] \; \int d\omega \;\;   \frac{f^-\left(\omega  \right)}{\epsilon - \omega + i0^+}  \,.
\end{align*}
\end{widetext}

\section{Pumped charge and spin current to lowest order in tunneling}
\label{appendix_lowest_order}

We make use of the rotational spin symmetry to rewrite the kinetic equations (\ref{kin_i}) and (\ref{kin_a}), see Appendix~\ref{appendix_general_trafo}.
An explicit calculation of the kernels yields
\begin{subequations}
\label{eq_master_1}
\begin{align}
\label{eq_master_spin}
\mathbf{0}
&= \mathbf{W}_p^{(\text{i},1)} \; \mathbf{P}_t^{(\text{i},0)} + \mathbf{v}_p^{(\text{i},1)} \;(\mathbf{S}_t^{(\text{i},0)}  \cdot \ep) \,,  \\
\label{eq_master_spin}
\mathbf{0}
&= (\mathbf{v}_\text{acc}^{(\text{i},1)} \cdot \mathbf{P}_t^{(\text{i},0)}) \ep - \frac{ \mathbf{S}_t^{(\text{i},0)} }{\tau^S_\text{rel}  }  
   +  B \;  \mathbf{S}^{(\text{i},0)} \times \ep  \,,
\end{align}
\end{subequations}
for the instantaneous limit and 
\begin{subequations}
\label{eq_master_2}
\begin{align}
\frac{d}{dt}\;\mathbf{P}_t^{(\text{i},0)}
&= \mathbf{W}_p^{(\text{i},1)} \; \mathbf{P}_t^{(\text{a},-1)} + \mathbf{v}_p^{(\text{i},1)} \;(\mathbf{S}_t^{(\text{a},-1)}  \cdot \ep) \,,  \\
\frac{d}{dt}\;\mathbf{S}_t^{(\text{i},0)}
&= (\mathbf{v}_\text{acc}^{(\text{i},1)} \cdot \mathbf{P}_t^{(\text{a},-1)}) \ep - \frac{ \mathbf{S}_t^{(\text{a},-1)} }{\tau^S_\text{rel}  }  
   +  B \; \mathbf{S}_t^{(\text{a},-1)} \times \ep  \,,
\end{align}
\end{subequations}
for the adiabatic correction.

The matrix $\mathbf{W}_p^{(\text{i},1)}$ and vector $\mathbf{v}_p^{(\text{i},1)}$ appearing in the kinetic equations for $\mathbf{P}$ are given by 
\begin{widetext}
\begin{align}
%
%
%
%
\mathbf{W}_p^{(\text{i},1)} \; & =    
\Gamma \left(
\begin{array}{ccc}
%
 -2 f^+(\epsilon) 
& f^{-}(\epsilon)
&0 \\	
%
2 f^+(\epsilon)
& - f^{-}(\epsilon)- f^+(\epsilon+U)
& 2  f^{-}(\epsilon+U)  \\	
%
0
&  f^+(\epsilon+U) 
& -2 f^{-}(\epsilon+U)  \\
\end{array}
\right) \\ 
%
%
%
%
%
%
%
%
\mathbf{v}_p^{(\text{i},1)} \; & =    2\, P
\Gamma_{\,\text{F}} \left(
\begin{array}{c}
 f^{-}(\epsilon) \\
 - f^{-}(\epsilon)  + f^+(\epsilon+U)  \\
-  f^+(\epsilon+U) 
\end{array}
\right) \, .
\end{align}
\end{widetext}
In the kinetic equations for the spin we have introduced 
\begin{align}
%
%
%
\label{eq_vacc}
\mathbf{v}_\text{acc}^{(\text{i},1)}
& =   P  \Gamma_{\,\text{F}} \left(
\begin{array}{c}
 f^+(\epsilon) \\
\frac{1}{2} \left[ - f^{-}(\epsilon) + f^+(\epsilon+U) \right]\\
-  f^-(\epsilon+U) 	
\end{array}
\right) 
\end{align}
as well as the spin relaxation time $\tau^S_\text{rel}$ defined by $1/\tau^S_\text{rel} = \Gamma \left[f^-(\epsilon)+ f^+(\epsilon+U)  \right] $ and the interaction-induced exchange field $B=  \frac{\Gamma_{\text{F}} \, P }{\pi} \; \pint d\omega  \left[ \frac{f^- (\omega)}{ \omega - \epsilon }  \;+\;   \frac{f^+ (\omega) }{ \omega - \epsilon - U } \right]$, where
$ \pint \, \, d\omega$ denotes Cauchy's principal value.
Here and in the following, we drop the energy dependence of the tunnel coupling $\Gamma_r(\omega) \equiv \Gamma_r$ and the polarization $P(\omega) \equiv P$. 
The generalization to an arbitrary energy dependence is straightforward.

As discussed in Appendix~\ref{appendix_general_trafo} the right hand side of the kinetic equation for the spin, Eq~(\ref{eq_master_spin}) has quite an intuitive interpretation.
The first term describes spin accumulation.
As a consequence of the spin rotational symmetry, the accumulation is along $\ep$.
The second term models spin relaxation.
For our model, the relaxation turns out to be isotropic, i.e., the relaxation times for the spin components parallel and perpendicular to the symmetry axis $\ep$ are identical and denoted by the same $\tau^S_\text{rel}$ in the following. 
The third term, describes a coherent rotation of the accumulated spin about an effective exchange field along the magnetization direction of the ferromagnetic lead.
Its magnitude $B$ depends on the dot level position $\epsilon$ and is, thus, tunable via the gate voltage.
It strongly depends on the Coulomb interaction. 
In fact,  when the energy dependence of $\gf$ and $P$ can be neglected, the exchange field vanishes in the absence of interaction $U=0$.
In addition to the predicted spin rotation,\cite{koenig_interaction_2003, braun_theory_2004} the exchange field leads to a splitting of the Kondo resonance,\cite{martinek_kondo_qd_2003} which has been experimentally confirmed recently.\cite{pasupathy_kondo_2004, hamaya_kondo_2007, hauptmann-2008}

From the kinetic equations~(\ref{eq_master_1}) and~(\ref{eq_master_2}), together with the normalization conditions $\mathbf{e}^\text{T}\,\mathbf{P}_t^{(\text{i},0)}=\,1$ and $ \mathbf{e}^\text{T}\;\mathbf{P}_t^{(\text{a},-1)}=\,0 $ we determine the probabilities and the spin $\mathbf{P}_t^{(\text{i},0)}$, $\mathbf{P}_t^{(\text{a},-1)}$, $\mathbf{S}_t^{(\text{i},0)}$, and $\mathbf{S}_t^{(\text{a},-1)}$, respectively.
We find that the instantaneous probabilities to lowest order in the tunnel coupling are the equilibrium values for the decoupled dot, thus the spin vanishes,
\begin{equation}
	\mathbf{S}_t^{(i,0)} = \mathbf{0} \, ,
\end{equation}
and the remaining occupation probabilities are simply given by Boltzmann factors,
%
%
\begin{equation}
\label{eq_P_i,0}
	\mathbf{P}_t^{(i,0)}  = \frac{1}{1+2e^{-\beta \epsilon} + e^{-\beta (2\epsilon+U)}}
	\left( \begin{array}{c} 1 \\	2e^{-\beta \epsilon} \\ e^{-\beta (2\epsilon+U)} \end{array} \right)
 \, ,
\end{equation}
where $\beta=1/(k_{\text{B}}T)$ is the inverse temperature. 
It follows that the instantaneous average occupation number $\langle n\rangle^{(i,0)}  = P_1^{(i,0)} +2P_\text{d}^{(i,0)} $ of the quantum dot in lowest order in $\Gamma$ is
%
%
\begin{equation}
\label{eq_n_i,0}
	\left\langle n \right\rangle^{(i,0)} 
	=\dfrac{2f^+(\epsilon)}{f^+(\epsilon)+f^-(\epsilon+U)} \, .
\end{equation}

The adiabatic corrections are given by
%
%
\begin{subequations}
\begin{align}
\label{P-a-1}
	\mathbf{P}_t^{(\text{a},-1)} &= -\; \tau^Q_\text{rel} \, 
	\frac{1}{1- \pol^2 \, \frac{\Gamma_{\rm{F}}^2}{\Gamma^2}} \;
	\dfrac{d}{dt} \mathbf{P}_t^{(\text{i},0)}\,,
\\
\label{S-a-1}
	\mathbf{S}_t^{(a,-1)}  &=  \frac{ \tau^S_\text{rel} }{2}\, \frac{P\frac{\Gamma_{\rm{F}}}{\Gamma}}{1-\pol^2 \, \frac{\Gamma_{\rm{F}}^2}{\Gamma^2} }\;  \dfrac{d}{d t}\langle n\rangle^{(i,0)}  \ep \, ,
\end{align}
\end{subequations}
where $\tau^Q_\text{rel}$ is the charge relaxation time given by $1/\tau^Q_\text{rel} = \Gamma \;  \left[ \,f^+(\epsilon) + f^-(\epsilon+U)  \, \right]  $.
The adiabatic correction to the probabilities  depends on the spin polarization of the ferromagnetic lead and on the tunnel-coupling strengths to both leads. Setting the polarization $P$ to zero we obtain the result for an N-dot-N structure.\cite{splettstoesser_adiabatic_2006}
The accumulated spin has only a component along the symmetry axis $\ep$.

We proceed in the same way for the charge and the spin currents flowing into the normal lead. We denote the charge current by  $I_{\text{N}}$  and the spin current by $\mathbf{J}_{\text{N}}$.
The instantaneous currents vanish, since the leads are kept at the same chemical potential (no bias voltage is applied.)
For the adiabatic corrections, we find 
\begin{subequations}
\begin{align}
\label{charge_current}
I_{\text{N}}^{(\text{a},0)}(t) 
&=\;-e\;\mathbf{v}^{\text{N}(\text{i},1)} \cdot \mathbf{P}_t^{(\text{a},-1)}   \\
\label{spin_current}
\mathbf{J}_{\text{N}}^{(\text{a},0)}(t)
&= \mathbf{S}_t^{(\text{a},-1)}  \;  \frac{\Gamma_{\text{N}} }{\Gamma}  \; \frac{1}{\tau^S_\text{rel}} 
\end{align}
\end{subequations}
where we defined the vector
\begin{align*}
%
%
\mathbf{v}^{\text{N}(\text{i},1)} \; 
& =    2 \;\Gamma_{\text{N}}
\left(
\begin{array}{c}
 f^+(\epsilon) \\
\frac{1}{2} \left[ - f^{-}(\epsilon) + f^+(\epsilon+U) \right]\\
-  f^-(\epsilon+U) 	
\end{array}
\right) \,.
\end{align*}
We remark that the accumulated spin does not enter the expression for the charge current and the probability vector does not enter the expression for the spin current.
This is a consequence of the fact that the currents are evaluated in the normal lead, which is not spin polarized.

The lowest-order contribution to the adiabatic current, given by Eqs.~(\ref{charge_current}) and (\ref{spin_current}), is linear in $\Omega$ and independent of $\Gamma$, in contrast to the dc current through a system with an applied transport voltage, which scales with $\Gamma$. Since $\Omega \ll \Gamma $, the pumped current goes to zero for vanishing tunnel coupling as it should. 

Plugging in the results for the probability vector and the spin, we finally obtain Eqs.~(\ref{ccur_a_0}) and (\ref{scur_a_0}).

\section{Pumped charge and spin current to first order in tunneling for pumping scheme B}
\label{appendix_first_order_B}

In pumping scheme B, the azimuth angle $\varphi(t)$ is the only time-dependent parameter.
As a consequence, $\mathbf{P}_t^{(\text{a}, -1)}$ and $\mathbf{S}_t^{(\text{a}, -1)}$ as given by Eqs.~(\ref{P-a-1}) and (\ref{S-a-1}) vanish, which leads to a vanishing charge and spin current to lowest order in the tunnel coupling, $I_{\text{N}}^{(\text{a},0)}(t) = 0$ and $\vec{J}_{\text{N}}^{(\text{a},0)}(t) = \mathbf{0}$.
It is, therefore, necessary to include the next order in the perturbation expansion in the tunnel-coupling strength.
The adiabatically pumped charge and spin currents to next order,
\begin{subequations}
\begin{align}
\label{charge_current_1}
I_{\text{N}}^{(\text{a},1)}(t) 
&=\;-e\;\mathbf{v}^{\text{N}(\text{i},1)} \cdot \mathbf{P}_t^{(\text{a},0)}   \\
\label{spin_current_1}
\mathbf{J}_{\text{N}}^{(\text{a},1)}(t)
&= \mathbf{S}_t^{(\text{a},0)} \;  \frac{\Gamma_{\text{N}} }{\Gamma}  \; \frac{1}{\tau^S_\text{rel}} 
\end{align}
\end{subequations}
depend on $\mathbf{P}_t^{(\text{a}, 0)}$ and $\mathbf{S}_t^{(\text{a}, 0)}$.
The latter are obtained from the kinetic equations to next-to-lowest order in the tunnel-coupling strength.

The kinetic equations expanded to next order in the tunnel coupling simplify if we use $\mathbf{P}_t^{(\text{a}, -1)}=\mathbf{0}$ and $\mathbf{S}_t^{(\text{i}, 0)}=\mathbf{S}_t^{(\text{a}, -1)}=\mathbf{0}$, resulting from Eqs.~(\ref{P-a-1}) and~(\ref{S-a-1}).
Furthermore, when solving the higher-order kinetic equations, it turns out that $\mathbf{P}_t^{(\text{i}, 1)}$ is constant in time and that $\mathbf{P}_t^{(\text{a}, 0)}=\mathbf{0}$.
This can be easily understood noting that the rotation of the ferromagnet's magnetization direction does not affect the probability distribution for empty, single, and double occupation.
It immediately follows that the pumped charge current is always zero.
Furthermore, we find that $\mathbf{S}_t^{(\text{i}, 1)}$ is always parallel to $\ep$.
This is consistent with the fact that in the instantaneous limit $\ep$ is a symmetry axis.

To simplify the presentation, we immediately make use of these results when writing down and solving the kinetic equations in the following.
To obtain the pumped spin current, we need $\mathbf{S}^{(\text{a},0)}$.
The latter is determined from the adiabatic correction to the kinetic equation for the spin,
\begin{align}
\label{kin-eq_sa0}
	\dfrac{d}{dt}\,\mathbf{S}_t^{(\text{i},1)}  
	= \mathbf{M}_p^{(\text{a},1)}\;\mathbf{P}_t^{(\text{i},0)} - \frac{ \mathbf{S}_t^{(\text{a},0)} }{\tau^S_\text{rel}  }  +  B \; \mathbf{S}_t^{(\text{a},0)} \times \ep \, .
\end{align}
All other terms that would formally appear in the expansion are vanishing, as mentioned above.
While $\tau^S_\text{rel}$ and $B$ are already known, and $\mathbf{M}_p^{(\text{a},1)}$ is straightforwardly constructed by applying the diagrammatic rules, the spin $\mathbf{S}_t^{(\text{i},1)}$ entering on the left hand side is still unknown.
To determine the latter, we write down the instantaneous kinetic equations for the spin in next-to-lowest order in the tunnel coupling, which immediately yields
%
\begin{align}
\frac{ \mathbf{S}_t^{(\text{i},1)} }{\tau^S_\text{rel}  } 
= \left( \mathbf{v}_\text{acc}^{(\text{i},1)} \cdot \mathbf{P}_t^{(\text{i},1)} + \mathbf{v}_\text{acc}^{(\text{i},2)}  \cdot \mathbf{P}_t^{(\text{i},0)}\right) \ep 
\end{align}
but depends on $\mathbf{P}_t^{(\text{i},1)}$.
To close the set of equations, we write down the instantaneous kinetic equations for the probabilities in next-to-lowest order in the tunnel coupling,
\begin{align}
\mathbf{0} = \mathbf{W}_p^{(\text{i},1)}\;\mathbf{P}_t^{(\text{i},1)}   + \; \mathbf{W}_p^{(\text{i},2)}\;\mathbf{P}_t^{(\text{i},0)}
+  \mathbf{v}_p^{(\text{i},1)}\; (\mathbf{S}_t^{(\text{i},1)} \cdot \ep) \, ,
\end{align}
and multiply it from the left with $\mathbf{q}^T= (0,1,2)$, since
$\mathbf{q}^T  \cdot \mathbf{W}_p^{(\text{i},1)} = 2\Gamma/(P\Gamma_\text{F}) (\mathbf{v}_\text{acc}^{(\text{i},1)})^T$ and 
$\mathbf{q}^T \cdot \mathbf{v}_p^{(\text{i},1)} = - 2P\Gamma_\text{F} /(\Gamma \tau^S_\text{rel}) $. 
This allows us to solve for $\mathbf{S}_t^{(\text{i},1)}$ and arrive at
\begin{align}
 \frac{ \mathbf{S}_t^{(\text{i},1)} }{\tau^S_\text{rel}  } 
=  \frac{  \mathbf{v}_\text{acc}^{(\text{i},2)} \cdot \mathbf{P}_t^{(\text{i},0)}		
	- \frac{P\Gamma_\text{F}}{2\Gamma} \mathbf{q}^T \mathbf{W}_p^{(\text{i},2)} \mathbf{P}_t^{(\text{i},0)}}
	{1-P^2\frac{\Gamma_\text{F}^2}{\Gamma^2}} \ep \, .
\end{align}
Now, $\mathbf{v}_\text{acc}^{(\text{i},2)}$ and $\mathbf{W}_p^{(\text{i},2)}$ can be evaluated with the help of the diagrammatic rules.

We remark that the kinetic equation~(\ref{kin-eq_sa0}) for the adiabatic correction $\mathbf{S}_t^{(\text{a},0)}$ to the spin contains two source terms.
One is given by the time derivative of the instantaneous spin $\mathbf{S}_t^{(\text{i},1)}$.
Since the latter is, for symmetry reasons, directed along $\ep$, the time derivative is along $ \dtep$.
The other source term is the adiabatic correction to the spin accumulation $\mathbf{M}_p^{(\text{a},1)} \;\mathbf{P}_t^{(\text{i},0)}$ which has components along $\ep \times \dtep$ and $ \dtep$.

Collecting everything, Eq.~(\ref{kin-eq_sa0}) reads
\begin{align}
\nonumber
	\mathbf{0}&=- \frac{1}{4}\, \frac{ \partial_\epsilon   \langle n \rangle^{(\text{i},0)}   }{\tau^Q_\text{rel} }  \left( \frac{\gf}{\g}  \;   \ep  \times \dtep - B \; \tau^S_\text{rel} \;   \dtep \right) \\
	&\quad - \frac{ \mathbf{S}_t^{(\text{a},0)} }{\tau^S_\text{rel}  }  +  B \; \mathbf{S}_t^{(\text{a},0)} \times \ep \, 
\label{kin-eq_sa0_1}
\end{align}
and we can eventually solve for the adiabatic correction of the spin and plug this into Eq.~(\ref{spin_current_1}) to get the pumped 
spin current as given in Eq.~(\ref{scur_a_1}).

\end{appendix}

\end{document}